\documentclass[12pt]{article}
\pdfoutput=1
\usepackage{array}
\usepackage{siunitx}

\usepackage{cite}
\usepackage{booktabs}
\usepackage[english]{babel}
\usepackage{amsmath,amssymb,amsbsy,amstext, amsthm, simplewick}
\usepackage{hyperref}
\usepackage{graphicx}
\usepackage{amsfonts}
\usepackage{amssymb}
\usepackage[small]{caption}
\usepackage{upgreek}
\usepackage[svgnames,dvipsnames,x11names,table]{xcolor}
\usepackage{multirow} 
\usepackage{geometry}
\usepackage[hang,flushmargin]{footmisc}
\usepackage{bm}
\usepackage{braket}
\usepackage{subcaption}
\usepackage{simpler-wick}
\usepackage{mathtools}
\usepackage{setspace}
\usepackage{cleveref}

\hypersetup{
    colorlinks=true,
    linkcolor={red!50!black},
    citecolor={blue!50!black},
    urlcolor={blue!80!black}
}

\usepackage{colortbl}

\usepackage[most]{tcolorbox}
\tcbset{colback=white, colframe=black,
        highlight math style= {enhanced, 
            colframe=red,colback=red!10!white,boxsep=0pt}
        }

\makeatletter
\newlength{\apb@width}
\newcommand{\autoparbox}[2][c]{\settowidth{\apb@width}{#2}\parbox[#1]{\apb@width}{#2}}

\makeatother

\definecolor{lightgray}{gray}{0.9}

\usepackage[framemethod=default]{mdframed}
\newmdenv[skipabove=7pt,
skipbelow=7pt,
rightline=false,
leftline=false,
topline=false,
bottomline=false,
backgroundcolor=gray!10,
linecolor=gray,
innerleftmargin=5pt,
innerrightmargin=5pt,
innertopmargin=5pt,
innerbottommargin=5pt,
leftmargin=0cm,
rightmargin=0cm,
linewidth=4pt]{eBox}

\crefname{table}{Table}{Tables}
\crefname{equation}{Eq.}{Eqs.}
\crefname{appendix}{App.}{Apps.}
\crefname{section}{Sec.}{Secs.}
\crefname{figure}{Fig.}{Figs.}

\numberwithin{equation}{section}

\def\beq{\begin{equation}}
\def\eeq{\end{equation}}

\def\bea{\begin{eqnarray}}
\def\eea{\end{eqnarray}}

\def\beq{\begin{equation}}
\def\eeq{\end{equation}}
\def\bea{\begin{eqnarray}}
\def\eea{\end{eqnarray}}

\def\x{{\vec x}}

\DeclareRobustCommand{\SkipTocEntry}[4]{}

\definecolor{colorTC}{rgb}{.2,.7,.2}

\definecolor{blue3}{RGB}{31, 119, 180}
\definecolor{red3}{RGB}{	214, 39, 40}
\definecolor{orange3}{RGB}{255, 127, 14}
\definecolor{green3}{RGB}{44, 160, 44}

\begin{document}

\begin{titlepage}
\setcounter{page}{1} \baselineskip=15.5pt 
\thispagestyle{empty}
$\quad$
\vskip 65 pt

\begin{center}
{\fontsize{18}{18} \bf The Phase of the BAO on Observable Scales}
\end{center}

\vskip 20pt
\begin{center}
\noindent
{\fontsize{12}{18}\selectfont  Daniel Green and Alexander K. Ridgway }
\end{center}

\begin{center}
\vskip 4pt
\textit{ {\small Department of Physics, University of California at San Diego,  La Jolla, CA 92093, USA}}

\end{center}

\vspace{0.4cm}
 \begin{center}{\bf Abstract}
 \end{center}
 
 \noindent The baryon acoustic oscillations (BAO) provide an important bridge between the early universe and the expansion history at late times.  While the BAO has primarily been used as a standard ruler, it also  encodes recombination era physics, as demonstrated by a recent measurement of the neutrino-induced phase shift in the BAO feature.  In principle, these measurements offer a novel window into physics at the time of baryon decoupling.  However, our analytic understanding of the BAO feature is limited, particularly for the range of Fourier modes measured in surveys.  As a result, it is unclear what the BAO phase teaches us about the early universe beyond what is already known from the cosmic microwave background (CMB).  In this paper, we provide a more complete (semi-)analytic treatment of the BAO on observationally relevant scales.  In particular, we compute corrections to the frequency and phase of the BAO feature that arise from higher order effects which occur in the tight coupling regime and during baryon decoupling.  The total phase shift we find is comparable to a few percent shift in the BAO scale (frequency) and thus relevant in current data.  Our results include an improved analytic calculation of the neutrino induced phase shift template that is in close agreement with the numerically determined template used in measurements of the CMB and BAO.  
 
\end{titlepage}

\restoregeometry

\begin{spacing}{1.5}
\newpage
\setcounter{tocdepth}{2}
\tableofcontents
\end{spacing}

\newpage

\section{Introduction}

Measurements of the Baryon Acoustic Oscillations (BAO)~\cite{Eisenstein:1997ik,Eisenstein:2005su} play an essential role in our understanding of the universe, connecting the physics  of recombination~\cite{Eisenstein:1997ik} and the expansion history at low redshift~\cite{Blake:2003rh}.  While the signal-to-noise of the measurement of the BAO~\cite{Beutler:2016ixs,Vargas-Magana:2016imr,Alam:2020sor} is much weaker than that of the cosmic microwave background (CMB)~\cite{Aghanim:2018eyx}, it is an invaluable tool for understanding the expansion history and breaking degeneracies.  For example, a future cosmological measurement of the sum of neutrino masses requires a DESI~\cite{Aghamousa:2016zmz} or Euclid~\cite{Amendola:2016saw} level BAO measurement of $\omega_m$ to accurately determine the amplitude of clustering~\cite{Zhen:2015yba,Allison:2015qca,Abazajian:2016yjj}.  Furthermore, with their ever-expanding size and depth, galaxy surveys are increasingly able to make competitive measurements without input from the CMB (see e.g.~\cite{Abbott:2017wau,Baumann:2018qnt,Beutler:2019ojk,Hill:2020osr,Philcox:2020vvt}).

Despite the importance of the BAO to observational cosmology, our analytic understanding of the linear BAO remains significantly less-developed than the CMB~\cite{Slepian:2015zra}.  Instead, the majority of theoretical work has been focused on important non-linear effects that uniquely impact the BAO measurements, such as nonlinear damping~\cite{Matsubara:2007wj,Crocce:2007dt,Sugiyama:2013gza,Porto:2013qua,Baldauf:2015xfa} and reconstruction~\cite{Eisenstein:2006nk,Padmanabhan:2008dd,Noh:2009bb,Tassev:2012hu,Zhu:2016sjc,Schmittfull:2017uhh}.  Nevertheless, measurements are at the point where additional BAO phase information can be measured~\cite{Baumann:2018qnt}.  Furthermore, the BAO peak location is degenerate with this phase information and could potentially bias the measurement of the expansion history~\cite{Baumann:2017gkg,Bernal:2020vbb}. This interplay between the BAO peak location and the phase is particularly interesting in light of the current tension in measurements of $H_0$~\cite{Aylor:2018drw,Knox:2019rjx}. 

The impact of cosmic neutrinos, and other light relics, presents a particularly compelling motivation to better understand corrections to the structure of acoustic peaks in the BAO~\cite{Green:2019glg}.  These particles travel faster than sound waves in the primordial plasma and source perturbations ahead of the acoustic horizon~\cite{Bashinsky:2003tk,Baumann:2015rya}.  On small scales, this effect appears as a constant shift in the phase of the acoustic peaks in the CMB and the BAO.  Furthermore, this shift is robust to non-linear evolution~\cite{Baumann:2017lmt} (like the BAO peak itself~\cite{Eisenstein:2006nj}).  Combining these insights, a first measurement of cosmic neutrinos was made in~\cite{Baumann:2018qnt} using data from BOSS DR12~\cite{Alam:2015mbd}. Yet, as with the CMB~\cite{Pan:2016zla}, there are important scale dependent corrections to the observed structure of the BAO in addition to the neutrino induced phase shift.  For example, it is well-known that the velocity overshoot~\cite{1970,Press:1980is} associated with the decoupling of photons and baryons implies a $\frac{\pi}{2}$ shift in phase between a Fourier mode describing a temperature fluctuations in the CMB and the corresponding BAO mode.  It is therefore natural to consider what additional contributions to this relative phase exist and what the measurement of the BAO could tell us about physics at the time of baryon decoupling.

Mirroring the description of the CMB, a simple qualitative description of the BAO can be described in terms of instantaneous decoupling of baryons and photons, and gives rise to an oscillatory contribution to the matter power spectrum~\cite{Hu:1995en},
\beq
\frac{\delta P(k)}{P(k)} \approx A_{\rm BAO} \,  c_s k  \,  \sin\left( k r_{\rm drift}\right) \ ,
\eeq
where $r_{\rm drift}$ is the sound horizon at the time of baryon decoupling and $A_{\rm BAO}$ is a constant.  The appearance of a sine (rather than a cosine) is a result of the velocity overshoot described above.  In this paper, we would like to understand the leading corrections to the phase of this oscillation on observables scales.  These can be described schematically as
\beq
\frac{\delta P(k)}{P(k)} \approx {\cal A}_{\rm BAO}  \,  c_s k  \,  \sin\left( k r_{\rm drift} +  \sum_i \varphi_i(k) \right) \ ,
\eeq
where each $\varphi_i(k)$ is a correction to the phase at leading order in some small parameter.  The CMB analogue of this calculation was performed in~\cite{Pan:2016zla} where they determined the contributions to the CMB peak locations in $\Lambda$CDM needed to match current observations.

Some of these phase corrections will be common to both the CMB and BAO.  These include the gravitational effects of matter and neutrinos prior to decoupling, as well as subleading corrections in the tight-coupling expansion.  These corrections change the photon distribution prior to recombination and thus directly impact the locations of acoustic peaks of the CMB and BAO.  In addition, the BAO phase is uniquely affected by baryon decoupling in a way that is distinct from the CMB.  We find four such contributions that correct the BAO phase, resulting from the photon evolution during the decoupling era and its impact on the baryon (and eventually matter) overdensity. We will see that these contributions are comparable in size to the neutrino phase shift or, alternatively, a few percent shift in the BAO scale $r_{\rm drift}$.  Current measurements are sensitive to sub-percent changes to the BAO scale~\cite{Beutler:2016ixs,Vargas-Magana:2016imr,Alam:2020sor}, and the neutrino induced phase alone~\cite{Baumann:2018qnt}, and thus should be sensitive to these additional phase shifts as well.

The broader motivation for this work is to understand how physics beyond $\Lambda$CDM might be uniquely encoded in the BAO.  However, such an exploration is challenging without first having a complete understanding of the phase in $\Lambda$CDM.  This is particularly pertinent to the BAO, in contrast to the CMB, since the data is typically reduced to a standard ruler which does not directly contain the phase information.  While this is valid in many models~\cite{Bernal:2020vbb}, the (semi)-analytic description of the BAO we describe in this paper should make a systematic exploration of the effects of new physics on the BAO phase more tractable.  

The paper is organized as follows: In section~\ref{sec:lin}, we review the equations of the linear evolution and the conventional derivation of the BAO signal.  In section~\ref{sec:tight}, we determine phase corrections associated with the tight coupling regime, including the effect of free-streaming neutrinos.   In section~\ref{sec:decouple}, we calculate the impact of physics at the time of baryon decoupling.  We combine these results in section~\ref{sec:BAO}, where we present our full description of the BAO phase and conclude in section~\ref{sec:conclusions}.   A few technical details of these calculations are relegated to appendices.

\section{Linear Theory of the BAO}\label{sec:lin}

The BAO feature in the matter power spectrum can be determined by solving the linear cosmological equations~\cite{Ma:1995ey}.  In this section, we give a simplified baseline treatment of the BAO which we later improve upon in sections \ref{sec:tight} and \ref{sec:decouple}.  We use Dodelson's~\cite{Dodelson:2003ft} conventions for the cosmological fluctuations throughout.

\subsection{Evolution Equations and Conventions}
\label{cosmological perturbations equation}

In this section, we enumerate the equations used to derive the BAO feature in the matter power spectrum.  We are interested in the evolution of the photons, baryons, dark matter and neutrinos which carry background energy densities $\bar \rho_\gamma$, $\bar \rho_{\rm b}$, $\bar \rho_{\rm dm}$, and $\bar \rho_\nu$. The metric can be expressed in conformal Newtonian gauge as
\beq
ds^2 = a(\eta)^2\left( - (1+2 \Psi(\x,\eta)) d\eta^2 + (1+ 2 \Phi(\x,\eta)) d\x^2 \right)
\eeq
where $a$ is the scale factor, $\Psi$ and $\Phi$ are the gravitational potentials, and $\eta$ is conformal time.  The fluctuations in baryons and dark matter are characterized by their overdensities $\delta_{\rm b}(\x, \eta)= \rho_{\rm b}(\x, \eta) / \bar \rho_{\rm b}(\eta)$ and $\delta(\x, \eta)= \rho_{\rm dm}(\x, \eta) / \bar \rho_{\rm dm}(\eta)$, and velocity potentials $\tilde{v}_{\rm b}$ and $\tilde{v}$.\footnote{The baryon and dark matter velocities are irrotational and can be written as divergences of the velocity potentials}  

The photon fluctuations are described by their temperature perturbation,
\begin{align}
f_{\gamma}(\vec{x},p,\hat{p},\eta) = \left[{\rm exp}\left(\frac{p}{T(\eta)(1+\Theta(\vec{x},\hat{p},\eta))}\right)-1\right]^{-1}, 
\end{align}
where $f_\gamma$ is the photon distribution function and $\Theta(\vec x, \hat{p}, \eta)$ represents the photon temperature fluctuation as a function of position, direction and time.  The neutrino temperature perturbation $\mathcal{N}(\vec{x},\hat{p},\eta)$ is defined similarly.  The temperature perturbations can be decomposed into multipole moments using a Legendre polynomial expansion: 
\begin{align}
\label{multipole exp}
\Theta_l \equiv \frac{1}{(-i)^l}\int_{-1}^{1}\frac{d\mu}{2}\mathcal{P}_l(\mu)\Theta(\mu) \rightarrow \Theta(\mu) = \sum_{l=0}^{\infty}(-i)^l(2l+1)P_l(\mu)\Theta_l
\end{align}
and similarly for the neutrinos.  
 
It is useful to express the equations of motion of the metric perturbations and cosmological components in momentum space.  Defining $\mu \equiv \hat{p}\cdot\hat{k}$, the relationship between the position and momentum space variables is given by
 \begin{align}
 \label{Theta fourier}
 \Theta(\vec{x}, \hat{p}, \eta) = \int \frac{d^3 k}{(2\pi)^3} e^{i \vec{k}\cdot\vec{x}} \Theta(\vec{k}, \mu, \eta)
 \end{align}
and similarly for the other components.  We are primarily interested in wave-numbers belonging to the observable BAO range which satisfy $0.1 h{\rm Mpc}^{-1} \leq k \leq 0.3 h{\rm Mpc}^{-1}$, where $k \equiv |\vec{k}|$.

The time evolution of the gravitational potentials is sourced by the first three moments of the photon and neutrino temperature perturbations, as well as by the baryon and dark matter overdensities and velocities.  Defining $v \equiv i\tilde{v}$ and $v_b \equiv i\tilde{v}_b$, the linear Einstein's equations are
\begin{align}
\label{Potential eqns}
&\dot{a} = H_0 \sqrt{\Omega_r + \Omega_m a}\\
\label{Poissons equation12}
&k^2 \Phi = 4 \pi G a^2 (\bar{\rho}_m \delta_m + 4\bar{\rho}_r \Theta_{r,0}+ \frac{3 a H}{k}(\bar{\rho}_m v_m + 4\bar{\rho}_r \Theta_{r,1}))\\
\label{gravity constraint equation}
&k^2\left( \Phi + \Psi\right) = -32 \pi G a^2 \bar{\rho}_r \Theta_{r,2}
\end{align}
where overhead dots represent derivatives with respect to $\eta$.  Note, the linear combinations of the matter and radiation perturbations that source the gravitational potentials are $\bar{\rho}_m \delta_m \equiv \bar{\rho}_{\rm dm} \delta + \bar{\rho}_b \delta_b$, $\bar{\rho}_m v_m \equiv \bar{\rho}_{\rm dm} v + \bar{\rho}_b v_b$, and $\bar{\rho}_r \Theta_r \equiv \bar{\rho}_\gamma \Theta + \bar{\rho}_\nu \mathcal{N}$.  We have also defined $\Omega_i \equiv \bar{\rho}_i/\rho_{cr}$, where $\rho_{cr}$ denotes the critical energy density.  With the exception of section \ref{Anisotropic stress section}, we set $\Theta_{r,2} = 0$ and freely make use of $\Phi=-\Psi$.

The dark matter only interacts gravitationally with the other components:
\begin{align}
\label{matter eqns}
\dot{\delta} + k v = -3\dot{\Phi},\ \dot{v} + \frac{\dot{a}}{a}v = k \Psi.
\end{align}
Baryons, on the other hand, experience gravitational and electromagnetic interactions with the other components.  Mass conservation implies
\begin{align}
\dot{\delta}_b + k v_b = -3\dot{\Phi}
\end{align}
and conservation of momentum yields
\begin{align}
\label{baryons EOM}
\tcboxmath{
\dot{v}_b + \frac{\dot{a}}{a}v_b = k \Psi + \frac{\dot{\tau}}{R}(v_b - 3\Theta_1)}
\end{align}
where we have defined
\begin{align}
\label{scattering thomp def}
\dot{\tau} = -\bar{n}_e \sigma_{\rm T}a,\ R=\frac{3\bar{\rho}_b}{4\bar{\rho}_\gamma}.
\end{align}
to parameterize the Thompson scattering of photons off free electrons.  In equation (\ref{scattering thomp def}), $\sigma_T$ is the Thompson scattering cross section and $\bar{n}_e$ is the background free electron number density.  

The equations describing the time evolution of the full photon and neutrino temperature perturbations are
\beq
\label{free stream neuts}
\begin{aligned}
\dot{\Theta} + ik\mu\Theta &= -\dot{\Phi} - ik\mu\Psi - \dot{\tau}(\Theta_0 - \Theta  -i\mu v_b - \frac{1}{2}P_2(\mu) \Pi)\cr
\dot{\mathcal{N}} + ik\mu \mathcal{N} &= -\dot{\Phi} - ik\mu\Psi.  
\end{aligned}
\eeq
where $\Pi\equiv \Theta_2 + \Theta^P_0 + \Theta^P_2$ and $\Theta^P$ denotes the temperature perturbation for photon polarizations.  Outside of section \ref{tight coupling}, which considers the effects of diffusion damping, we set $\Pi = 0$.  Using (\ref{multipole exp}), (\ref{free stream neuts}) can be broken into equations describing the first few multipole moments:
\begin{align}
\label{rad eqns}
&\ \ \ \ \ \ \ \dot{\mathcal{N}}_0 + k\mathcal{N}_1 = -\dot{\Phi},\ \dot{\mathcal{N}}_1 - \frac{k}{3}(\mathcal{N}_0 - 2\mathcal{N}_2) = \frac{k}{3}\Psi
\\
\label{photon EOM moments multi}
&\dot{\Theta}_0 + k \Theta_1 = -\dot{\Phi},\ \dot{\Theta}_1 - \frac{k}{3}(\Theta_0 - 2\Theta_2) =  \frac{k}{3}\Psi + \frac{\dot{\tau}}{3}(3 \Theta_1 - v_b).
\end{align}
It is often convenient to use the $k\Theta_1 =- \dot \Theta_0- \dot \Phi$ in the dipole equation to get 
\begin{align} \label{eq:box_theta}
\tcboxmath{
(\partial_\eta^2 + \frac{k^2}{3})(\Theta_0 + \Phi) = -\frac{\dot{\tau} k}{3}(3 \Theta_1 - v_b) + \frac{2}{3}k^2 \Theta_2 -\frac{k^2 }{3}\left(\Psi-\Phi\right)}
\end{align}
The LHS takes the form of a wave-equation, as expected.  Outside of sections \ref{tight coupling} and \ref{sec:quadrupole}, where we consider corrections to the tight coupling approximation and free-streaming effects near decoupling, we set $\Theta_2 = 0$.  In the tight coupling regime, it is useful to eliminate $v_b$ using (\ref{baryons EOM}).

In this paper, the initial conditions of the cosmological variables are expressed in units of the primordial adiabatic fluctuation $\zeta \equiv \Phi  -ik_i\delta T^{0}_i H/(k^2(\rho + P))$ as $\tau \to 0$.  Since we are interested in the linear impact of baryons on the matter transfer function, which is independent of the statistics of the primordial fluctuations, we are free to set $\zeta = 1$ throughout.

\subsection{Baseline Derivation of the BAO}
\label{baseline sec}
We begin by giving a simplified derivation of the BAO feature in the matter overdensity, following~\cite{Hu:1995en}.  From the start of radiation domination to the era of decoupling, the baryons and photons are tightly coupled and oscillate in phase with each other, i.e. $v_b = 3\Theta_1 + O(k/\dot{\tau})$.  The pressure from photon scatterings prevents the gravitational collapse of baryons and the amplitude of the baryon perturbation is comparable to that of the photons.  Over time, the expansion of the universe causes the photon-baryon scattering rate to decrease.  The baryons eventually decouple from the photons and undergo gravitational collapse.  The oscillation frequency of the photon-baryon system depends on the wavenumber $k$, meaning different baryon modes decouple at different phases in their oscillations.  The initial conditions for the gravitational growth of the baryons then have an oscillatory shape which gives rise to the BAO.

It is useful to work backwards and first consider the evolution of the baryons after they decouple from the photons.  Combining Poisson's equation (\ref{Poissons equation12}) with the evolution equations for the dark matter (\ref{matter eqns}) and baryons (\ref{baryons EOM}) gives a closed set of equations describing the full matter perturbation:
\begin{align}
\label{total matter EOM}
\dot{\delta}_m + k v_m =0,\ \dot{v}_m + \frac{\dot{a}}{a}v_m = -k\Phi,\ k^2\Phi = \frac{3H_0^2}{2a}\Omega_m\delta_m.
\end{align}
Since $k \gg aH$ near decoupling, $\delta_m$ is well approximated by the large $k$ solutions first obtained by Meszaros~\cite{Meszaros:1974tb},
\begin{align}
\label{matter sol full}
\delta_m(k,y) = C_{1}(k)D_1(y) + C_{2}(k)D_2(y)
\end{align}
where
\begin{align}
D_1(y) = y+2/3,\ D_2(y) = D_1(y){\rm ln}\left(\frac{\sqrt{1+y}+1}{\sqrt{1+y}-1} \right)-2\sqrt{1+y},\ y = a/a_{\rm eq}.
\end{align}
In the limit of late times, the growing solution $D_1$ scales as $D_1(y)\sim y$ while the decaying one $D_2$ scales as $D_2(y) \sim 1/y^{3/2}$.  Therefore,  the resulting matter overdensity at low-redshift is given by $\delta_m(k,y) \simeq C_1(k) D_1(y)$.

Poisson's equation in (\ref{total matter EOM}) implies that once the growing mode dominates, the position dependence of the gravitational potential is proportional to the Fourier transform of $C_1(k)/k^2$.  This encodes the location of the gravitational potential wells that both dark matter and baryons fall into and eventually reside in at late times.  The $C_i(k)$ depend on the initial locations of the dark matter and baryons as they begin to undergo gravitational growth.  They are fixed by matching $\delta_m$ and its derivative to the expression for $\delta_m$ valid before the baryons fully decouple from the photons.  The matching conditions are
\beq
\begin{aligned}
&C_1(k)D_1(y(\eta_g)) + C_2(k)D_2(y(\eta_g)) = \frac{\Omega_b}{\Omega_m} \delta_b(\eta_g) + \frac{\Omega_{\rm cdm}}{\Omega_m} \delta(\eta_g),\cr &C_1(k)\dot{D}_1(y(\eta_g)) + C_2(k)\dot{D}_2(y(\eta_g)) = \frac{\Omega_b}{\Omega_m} \dot \delta_b(\eta_g) + \frac{\Omega_{\rm cdm}}{\Omega_m}\dot \delta(\eta_g).
\end{aligned}
\eeq 
We choose $\eta_g$ to be some when the gravitational force on baryons dominates over photon-baryon scatterings ($\dot{\tau} \ll k \Phi$) and the dark matter overdensity is still the largest source of gravity ($\delta \gg \delta_b$).  The growing mode coefficient can be split into terms coming from the dark matter and baryons, $C_{1} \equiv C_{\rm cdm} + C_{\rm BAO}$.\footnote{Restoring units of $\zeta$, the BAO feature in the matter power spectrum is related to $C_{\rm BAO}$ through $P^{\rm BAO}_{\delta}(k,a) = 2D_1(a)^2C_{\rm cdm}(k)C_{\rm BAO}(k)P_{\zeta}(k)$.}  Focusing on the baryonic piece, the matching condition gives
\begin{align}
\label{matter growing mode coeff}
C_{\rm BAO}(k)&=  \frac{\Omega_{b}}{\Omega_m}\frac{D_2}{D_1\dot{D}_2 - D_2\dot{D}_1}(k v_b + \frac{\dot{D}_2}{D_2}\delta_b )\simeq -\frac{2k}{5\dot{y}}\frac{\Omega_{b}}{\Omega_m}(v_b -\frac{3aH}{2k}\delta_b )
\end{align}
where all functions of time in (\ref{matter growing mode coeff}) are evaluated at $\eta_g$.  In the last equality, we approximated $y(\eta_g)\gg 1$ and replaced the $D_i$ with their asymptotic expressions.  

Since $k \gg aH(\eta_g)$, the term proportional to $\delta_b$ can be neglected, giving
\begin{align}
\label{simplified vb grow}
C_{\rm BAO}(k) \simeq -\frac{2k}{5\dot{y}(\eta_g)}\frac{\Omega_{b}}{\Omega_m}v_b(k,\eta_g).
\end{align}
The leading behavior of the BAO profile is then given by the scale dependence of $v_b(k,\eta_g)$.  The fact that the BAO is fixed by the scale dependence of $v_b$ and not $\delta_b$ is the well known ``velocity overshoot" effect~\cite{1970,Press:1980is}.  Equations (\ref{matter sol full}) and (\ref{simplified vb grow}) then give the BAO feature in the total matter overdensity. 

To determine $v_b(k,\eta_g)$ and evaluate (\ref{simplified vb grow}), one simply integrates (\ref{baryons EOM}): 
\begin{align}
\label{v_b full 2}
v_b(k,\eta_g) =  - \frac{3}{a(\eta_g)}\int_{-\infty}^{\eta_g} d\eta a(\eta) g_b(\eta)\Theta_1(k,\eta) + I_{\Phi}(k,\eta_g).
\end{align}
We have defined the {\it baryon visibility function}~\cite{Hu:1995en} $g_b(\eta)$ through
\begin{align}
\label{definition of b optical depth}
&\tau_b(\eta) \equiv \int^\eta d\eta' \frac{\dot{\tau}}{R},\ g_b \equiv \dot{\tau}_be^{-\tau_b}     
\end{align}
and written the contribution from gravity as
\begin{align}
\label{integral from gravity}
&I_{\Phi}(k,\eta) \equiv -\frac{k}{a(\eta)}\int_{-\infty}^{\eta} d\eta' a(\eta')e^{\tau_b(\eta) - \tau_b(\eta')} \Phi(\eta').
\end{align}
The first term in (\ref{v_b full 2}) gives the cumulative effect of the photon driving force on $v_b$ and is responsible for the oscillatory shape of the BAO.  The contribution from $I_\Phi$ has no oscillatory information and only shifts the average value of the BAO feature.  

The evolution of the tightly coupled photons can be obtained by using (\ref{baryons EOM}) to eliminate $(3\Theta_1 - v_b)$ from (\ref{eq:box_theta}), yielding
\begin{align}
\label{photon monopole sol}
    (\partial_\eta^2 + k^2 c_s^2)(\Theta_0 + \Phi) = \frac{k^2}{3}(1 + \frac{1}{1+R})\Phi,
\end{align}
where $c_s = 1/\sqrt{3(1+R)}$ gives the speed of sound of the photon-baryon fluid.  In deriving (\ref{photon monopole sol}), we have only kept leading order terms in $k/\dot{\tau}$ (e.g.~$v_b = 3 \Theta_1+ {\cal O}(k/\dot{\tau} )$) and dropped the drag term proportional to $\dot{a}/a$.  Near decoupling, the gravitational source in (\ref{photon monopole sol}) is negligible and the photon perturbations are given by
\begin{align}
\label{theta0 near etastar}
\Theta_0(k,\eta) + \Phi(k,\eta)  &\simeq  - {\rm cos}(kr_s(\eta)) \implies \Theta_1(k,\eta) \simeq  -c_s(\eta){\rm sin}(kr_s(\eta)) 
\end{align}
where $r_s(\eta) = \int^\eta d\eta'c_s(\eta')$ is the sound horizon of the photon-baryon fluid.  The minus signs in (\ref{theta0 near etastar}) are due to the evolution of the photon perturbation during radiation domination (see Appendix \ref{rad matt matching app} for details).  In (\ref{theta0 near etastar}), we have only kept the leading contribution from this effect.

Equation (\ref{photon monopole sol}) is the leading expression for $\Theta_1$ near decoupling and can be combined with (\ref{v_b full 2}) to obtain $v_b(k,\eta_g)$.  The baryon visibility function $g_b(\eta)$ represents the probability for an individual baryon to last scatter with the photons at $\eta$ and can be approximated as a Gaussian:
\begin{align}
\label{Gaussian approx}
g_b(\eta) \simeq -\frac{1}{\sigma\sqrt{\pi}} e^{-\frac{(\eta - \eta_*)^2}{\sigma^2}}.
\end{align}
Since $\eta_g$ is chosen after the baryons have fully decoupled from the photons, i.e. $g_b(\eta_g)\simeq 0$, we can take the upper bound of the integral in (\ref{v_b full 2}) to infinity.  Using the fact that (\ref{Gaussian approx}) is peaked at $\eta_*$, the integrand can be simplified by setting $kr_s(\eta) \simeq kr_s(\eta_*) + kc_s(\eta_*)(\eta-\eta_*)$ and $a(\eta) \simeq a(\eta_*)$.  

Evaluating  (\ref{v_b full 2}) with these approximations gives our leading expression for the BAO profile
\begin{align}
\label{zeroth order BAO}
C_{\rm BAO}(k) = -\frac{2k}{5\dot{y}(\eta_g)}\frac{\Omega_b}{\Omega_m}\left[-\mathcal{A}(k){\rm sin}(k r_s(\eta_*))+I_\Phi(k,\eta_g)\right]
\end{align}
where the amplitude is given by
\begin{align}
\label{simple BAO amplitude}
\mathcal{A}(k) \equiv 3\frac{a(\eta_*)}{a(\eta_g)}c_s(\eta_*)e^{-\left(\frac{kc_s(\eta_*)\sigma}{2}\right)^2}.
\end{align}
As already mentioned, the appearance of the sine in (\ref{zeroth order BAO}) is a consequence of the velocity overshoot effect, and the fact that the baryon velocity is determined by the photon dipole (\ref{simplified vb grow}).  The exponential in (\ref{simple BAO amplitude}) suppresses modes that oscillate rapidly within the support of $g_b$, i.e. $kc_s(\eta_*) \gg 1/\sigma$.

In deriving (\ref{zeroth order BAO}), we neglected several important effects that change the evolution of the tightly coupled photons.  In addition, the treatment of the photon-baryon dynamics near decoupling was greatly simplified.  Corrections to the tightly coupled photons and the physics near decoupling impact the BAO profile and are considered in sections \ref{sec:tight} and \ref{sec:decouple}.

\section{Corrections to Tightly Coupled Photons}
\label{sec:tight}

The expression for the tightly coupled photons in (\ref{theta0 near etastar}) neglects gravitational driving during matter domination, the gravitational influence of neutrino anisotropic stresses and higher order corrections to the tight coupling approximation.  The contributions from these effects change the phase and amplitude of $\Theta_1$ and the driving force the photons exert on the baryons.  In this section, we compute the leading corrections from each of these effects on the photon evolution.  We first treat the effects separately and then combine terms in section \ref{combine improved tight coupling}.

\subsection{Gravity Driving During Matter Domination}
\label{gravity driving during matter dom}

In deriving our original expression for the photons (\ref{theta0 near etastar}) during matter domination, we neglected the gravitational source on the right hand side of (\ref{photon monopole sol}) and only kept the homogeneous solution.  As we will show, gravitational driving during matter domination changes the zero point of the photon oscillations and shifts their phase and amplitude.  

The leading contribution from gravity driving during matter domination is obtained by convolving the right hand side of (\ref{photon monopole sol}) with the Green's function for the tightly coupled photons,
\begin{align}
\label{Green's function for tightly coupled photons}
G_\Theta(\eta,\eta') = -\frac{1}{c_s k}({\rm cos}(kr_s(\eta)){\rm sin}(kr_s(\eta')) - {\rm cos}(kr_s(\eta')){\rm sin}(kr_s(\eta)))\Theta(\eta-\eta'),
\end{align}
from the time of matter-radiation equality $\eta_{\rm eq}$ defined through $\bar{\rho}_r(\eta_{\rm eq}) = \bar{\rho}_m(\eta_{\rm eq})$, to some time $\eta$ before decoupling.  We find 
\begin{align}
\label{Theta0 with potential}
    \Theta_0 (k,\eta) + \Phi(k,\eta) = -(1 - \alpha(k,\eta)){\rm cos}(k r_s(\eta)) +  \beta(k,\eta){\rm sin}(k r_s(\eta))
\end{align}
where 
\beq
\begin{aligned}
\label{pieces from Greens fct}
\alpha(k,\eta) &= -2k\int_{r_s(\eta_{\rm eq})}^{r_s(\eta)}dx{\rm sin}(kx)\Phi(k,\sqrt{3}x)\cr \beta(k,\eta) &= 2k\int_{r_s(\eta_{\rm eq})}^{r_s(\eta)}dx{\rm cos}(kx)\Phi(k,\sqrt{3}x).
\end{aligned}
\eeq
In writing (\ref{pieces from Greens fct}), we changed integration variables to $x = r_s(\eta)$ and approximated $1 + \frac{1}{1+R}\simeq 2$.

To evaluate (\ref{pieces from Greens fct}), we need an expression for the gravitational potential $\Phi$ valid in the range $\eta_{\rm eq} < \eta < \eta_*$.  During this time, the dark matter overdensity gives the largest contribution to $\Phi$ since $\delta_b \sim \Theta_0 \ll \delta$.  The equations describing $\delta$ and $\Phi$ are identical to (\ref{total matter EOM}) with $\delta_m$ and $v_m$ replaced with $\delta$ and $v$.  One can approximate $k \gg aH$ during matter domination, which means $\delta(k,\eta) = \delta_1(k)D_1(y) + \delta_2(k)D_2(y)$.  Neglecting the decaying mode $D_2$, the gravitational potential takes the form 
\begin{align}
\label{matt dom potential 1}
\Phi(k,y) = \frac{3}{2k^2}\frac{H_0^2}{\Omega_r}\Omega_m\Omega_{\rm dm}\delta_1(k)(1+\frac{2}{3 y}) \equiv \Phi_k(1+\frac{2}{3 y}).
\end{align}
The scale dependence of the gravitational potential $\Phi_k$ can be obtained by fitting (\ref{matt dom potential 1}) to the numerical solution of $\Phi$ (e.g. using matter transfer function output from a Boltzmann code like CLASS~\cite{Blas:2011rf} or a fitting function like BBKS~\cite{Bardeen:1985tr}).  

The fact that the scale and time dependence of $\Phi(k,y)$ factorizes simplifies (\ref{pieces from Greens fct}).  Working in the limit $kr_s(\eta_{\rm eq}) \gg 1$, and using $y = a(\eta)\Omega_m /\Omega_r$ and $a(\eta) = \frac{1}{4}\eta^2 H_0^2 \Omega_m + \eta H_0 \sqrt{\Omega_r}$, we find
\beq
\begin{aligned}
\label{phi integration theta0}
&\alpha(k,\eta) = {\rm cos}(kr_s(\eta))f_\Phi(k,\eta) - {\rm cos}(kr_s(\eta_{\rm eq}))f_\Phi(k,\eta_{\rm eq})\cr
&\beta(k,\eta) = {\rm sin}(kr_s(\eta))f_\Phi(k,\eta) - {\rm sin}(kr_s(\eta_{\rm eq}))f_\Phi(k,\eta_{\rm eq})\cr
&\tilde{x} = \frac{4}{H_0\Omega_m}\sqrt{\frac{\Omega_r}{3}},\ f_\Phi(k,\eta) = 2\Phi_k\left(1+\frac{\tilde{x}}{6}\left(\frac{1}{r_s(\eta)}-\frac{1}{r_s(\eta) + \tilde{x}} \right)\right).
\end{aligned}
\eeq
Inserting (\ref{phi integration theta0}) into (\ref{Theta0 with potential}) gives 
\begin{align}
\label{mat dom gravity shift photons}
&\Theta_0 (k,\eta) + \Phi(k,\eta) = f_\Phi(k,\eta) - (1 - A_{\Phi}(k)){\rm cos}(k r_s(\eta)) + B_{\Phi}(k){\rm sin}(k r_s(\eta))
\end{align}
where we have defined
\begin{align}
A_{\Phi}(k) = -{\rm cos}(kr_s(\eta_{\rm eq}))f_\Phi(\eta_{\rm eq}),\ B_{\Phi}= -{\rm sin}(kr_s(\eta_{\rm eq}))f_\Phi(\eta_{\rm eq})).
\end{align}
Equation (\ref{mat dom gravity shift photons}) implies gravitational driving during matter domination shifts the zero point of the photon oscillations by $f_\Phi$, and induces the phase shift
\begin{align}
\label{phase shift grav driving}
\phi_{\Phi} \simeq {\rm sin}^{-1}(B_{\Phi}).
\end{align}

The phase shift is then proportional to the transfer function of the gravitational potential $\Phi_k$.  It is well known that for modes entering the horizon before matter-radiation equality, the transfer function decreases with increasing $k$.  The observable BAO modes satisfy $k > aH(\eta_{\rm eq})$, which means the BAO phase shift resulting from $\phi_\Phi$ is largest for modes at the lower end of the observable range.

\subsection{Neutrino Anisotropic stresses}
\label{Anisotropic stress section}

In this section, we compute the effects of neutrino anisotropic stress on the evolution of the photons.  Unlike the tightly coupled photons, neutrinos free stream at nearly the speed of light in vacuum during radiation domination.  This means their anisotropic stress, parametrized by $\mathcal{N}_2$, is not suppressed relative to $\mathcal{N}_0$ or $\mathcal{N}_1$.  Since neutrinos travel faster than the speed of sound of the photon-baryon fluid, they source a unique contribution to the gravitational potential that is correlated at distances larger than $r_s(\eta)$.  This induces a phase shift in the photon oscillations which changes the peak locations of both the CMB \cite{Bashinsky:2003tk,Baumann:2015rya} and BAO \cite{Baumann:2017lmt}.

The gravitational effects of free streaming neutrinos on photons can be computed similarly to how the gravitational effects of matter were evaluated in section \ref{gravity driving during matter dom}.  The difference here is that since we are including a finite $\mathcal{N}_2$, the constraint equation (\ref{gravity constraint equation}) implies $\Phi \neq -\Psi$.  The gravitational source on the right hand side of (\ref{photon monopole sol}) now consists of the combination\footnote{The relationship between the $\Phi_{\pm}$ variables defined here and in \cite{Bashinsky:2003tk,Baumann:2015rya} are $\Phi_{\pm} \rightarrow \mp \Phi_{\mp}$.} $\Phi_- \equiv \Phi - \Psi$.  The leading contribution from neutrino anisotropic stress can be written 
\beq
\begin{aligned}
\label{fs for phase}
&\Theta_0 (k,\eta) + \Phi(k,\eta) =  -(1 - A_{\nu}(k,\eta)){\rm cos}(k r_s(\eta)) +  B_{\nu}(k,\eta){\rm sin}(k r_s(\eta))\cr
&A_{\nu}(\eta) = -kc_s\int_0^\eta d\eta' {\rm sin}kr_s(\eta')\Phi^{(1)}_-(\eta'),\ B_{\nu}(\eta) = kc_s\int_0^\eta d\eta' {\rm cos}kr_s(\eta')\Phi^{(1)}_-(\eta')
\end{aligned}
\eeq
where $\Phi^{(1)}_-$ denotes the leading correction to $\Phi_-$ from the gravitational effects of $\mathcal{N}_2$.

According to equation (\ref{gravity constraint equation}), the symmetric combination $\Phi_+ \equiv \Phi + \Psi$ is sourced by the neutrino anisotropic stress.  To determine $\mathcal{N}_2$, it is useful to first solve (\ref{free stream neuts}) for the full neutrino temperature perturbation, 
\begin{align}
\label{neutrino temp ev}
{\mathcal N}(\mu,\eta) = {\mathcal N}_{\rm in}e^{-ik\mu\eta} - \int_0^\eta d\eta' e^{-ik\mu(\eta-\eta')}(\dot{\Phi} + ik\mu\Psi),
\end{align}
and to break (\ref{neutrino temp ev}) into multipole moments using the plane wave expansion
\begin{align}
\label{Neutrino multipoles}
&e^{-ik\mu \eta} = \sum_{l=0}^{\infty}(-i)^l(2l+1)P_l(\mu)j_l(k\eta)
\end{align}
and (\ref{multipole exp}) for neutrinos.  Using (\ref{neutrino temp ev}) and (\ref{Neutrino multipoles}) and projecting out the quadrupole component, we find
\begin{align}
\label{neutquad}
{\mathcal N}_2(x) = ({\mathcal N}_{\rm in} + \Phi_{\rm in}) j_2(c_s^{-1}x) - c_s^{-1} \int_0^x dx' j_2'(c_s^{-1}(x-x'))\Phi_-(x').
\end{align}
where $x\equiv k c_s \eta$.

The largest gravitational interactions between free streaming neutrinos and photons occur during radiation domination.  This can be understood heuristically by considering the evolution of initial spikes in the photon-baryon fluid and neutrino distribution.  Since the neutrinos travel faster than the tightly coupled photons, the distance between the neutrino spike and the photon-baryon one gets larger as time goes on and the strength of their gravitational interaction decreases.  To a good approximation, the effect of free streaming neutrinos on the photons can be computed using expressions valid during radiation domination.  Only a small error is accumulated by using these expressions during matter domination, because by this time the gravitational effects of free-streaming neutrinos on photons has become small.

The gravitational effect of $\mathcal{N}_2$ can be derived by combining (\ref{rad eqns}) and (\ref{photon EOM moments multi}) with (\ref{Poissons equation12}) and (\ref{gravity constraint equation}).  Using $a(\eta) \sim \eta$ and the fact that the modes of interest are well inside the horizon during radiation domination (i.e. $k\eta \gg 1$), we find
\begin{align}
\label{neutrinos eqns}
&\Phi''_- + \frac{4}{x}\Phi'_- + \Phi_- = -S(\Phi_+),\ S(\Phi_+) \equiv \Phi''_+ +\frac{2}{x}\Phi'_+ + 3 \Phi_+,\ \Phi_+(x) = -\frac{4\epsilon_\nu}{x^2}{\mathcal N}_2(x).
\end{align}
Equation (\ref{neutrinos eqns}) can be computed perturbatively in $\epsilon_\nu$.  The leading correction to $\Phi_-$ is given by
\beq
\begin{aligned}
\label{potentials m}
&\Phi^{(1)}_-(x) = -\frac{4\epsilon_\nu}{15}\frac{{\rm sin}x-x{\rm cos}x}{x^3}   - \int_0^x dx' G(x,x')S(\Phi_+(x'))\cr
&G(x,x') = \Theta(x-x')\frac{x'}{x^3}((x'-x){\rm cos}(x'-x)-(1+xx'){\rm sin}(x'-x)),
\end{aligned}
\eeq
where the first term is due to the change in initial conditions from neutrino anisotropic stress (see Appendix \ref{initial conditions}).  

We are interested in evaluating (\ref{fs for phase}) at $\eta_*$, the peak of the baryon visibility function.  It is useful to write
\begin{align}
\label{f nu asymptotic}
B_{\nu}(k,\eta_*) =  \int_0^\infty dx\,{\rm cos}x\,\Phi_-^{(1)}(x) - \int_{k r_s(\eta_*)}^\infty  dx\,{\rm cos}x\,\Phi^{(1)}_-(x)
\end{align}
and similarly for $A_\nu$.  The first term in (\ref{f nu asymptotic}) is scale-independent and can be evaluated numerically, while
the remaining integral is scale dependent.  Since $kr_s(\eta_*) \gg 1$, the second integral can be computed perturbatively in $1/kr_s(\eta_*)$.

In the limit $x \rightarrow \infty$, the source function $S(\Phi_+(x))$ in (\ref{potentials m}) decays to zero.  This can be seen directly from the expression for $\mathcal{N}_2$ in (\ref{neutquad}) since the factors of spherical Bessel functions and $\Phi_-$ vanish in this limit. The physical reason the source vanishes is because the spatial separation between the free streaming neutrinos and sound horizon gets large as $x\rightarrow \infty$ and, as a result, the gravitational source function decays to zero.

Since $S(x)\rightarrow 0$ in the limit of large $x$, the $x'$ integral in (\ref{potentials m}) can be approximated by extending the upper bound from $x$ to $\infty$.  The dependence on $x$ and $x'$ then factorizes and the integrations over $x'$ can be performed numerically.  The $O(\epsilon_\nu)$ integrand can be obtained by first plugging the zeroth order solution of $\Phi_-$ (i.e. the homogeneous solution of (\ref{neutrinos eqns})) into (\ref{neutquad}) to find the zeroth order expression for $\mathcal{N}_2$.   Combining the result with (\ref{neutrinos eqns}) then gives $S(\Phi_+)$ to $O(\epsilon_\nu)$.  After performing the $x'$ integrals numerically, equation (\ref{potentials m}) yields
\begin{align}
\label{large x phi-1}
\Phi^{(1)}_-(x \gg 1) \simeq -\frac{4\epsilon_\nu}{15}\frac{{\rm sin}x-x{\rm cos}x}{x^3}   + \frac{{\rm cos}x}{x^3}(\kappa_1-x\kappa_2) + \frac{{\rm sin}x}{x^3}(\kappa_2 + x\kappa_1)
\end{align}
where
\beq
\begin{aligned}
\kappa_1 &\equiv - \int_0^\infty dx' (x'^2{\rm cos}x' - x' {\rm sin}x')S(x') \simeq 2.3 \epsilon_\nu\cr
\kappa_2 &\equiv - \int_0^\infty dx' (x'^2{\rm sin}x' + x' {\rm cos}x')S(x') \simeq -4.8 \epsilon_\nu.
\end{aligned}
\eeq
The scale dependent integral in (\ref{f nu asymptotic}) can be computed analytically using (\ref{large x phi-1}).  The gravitational influence of neutrino anisotropic stress is then given by
\beq
\begin{aligned}
\label{f nus pieces}
&A_{\nu}(k,\eta_*) = 0.44\epsilon_\nu + \frac{\kappa_1}{2 kr_s}+\frac{15\kappa_1{\rm sin}2kr_s+(15 \kappa_2 -4\epsilon_\nu)(1-{\rm cos}2kr_s)}{60(kr_s)^2}\cr 
&B_{\nu}(k,\eta_*) = 0.58 \epsilon_\nu +  \frac{1}{kr_s}(\frac{\kappa_2}{2}-\frac{2\epsilon_\nu}{15}) - \frac{15 \kappa_1(1+{\rm cos}2kr_s)+(15\kappa_2-4\epsilon_\nu){\rm sin}2kr_s}{60(kr_s)^2}
\end{aligned}
\eeq
where all factors of $r_s$ in (\ref{f nus pieces}) are evaluated at $\eta_*$. 

\begin{figure}[h!]
\centering
\includegraphics[width=5in]{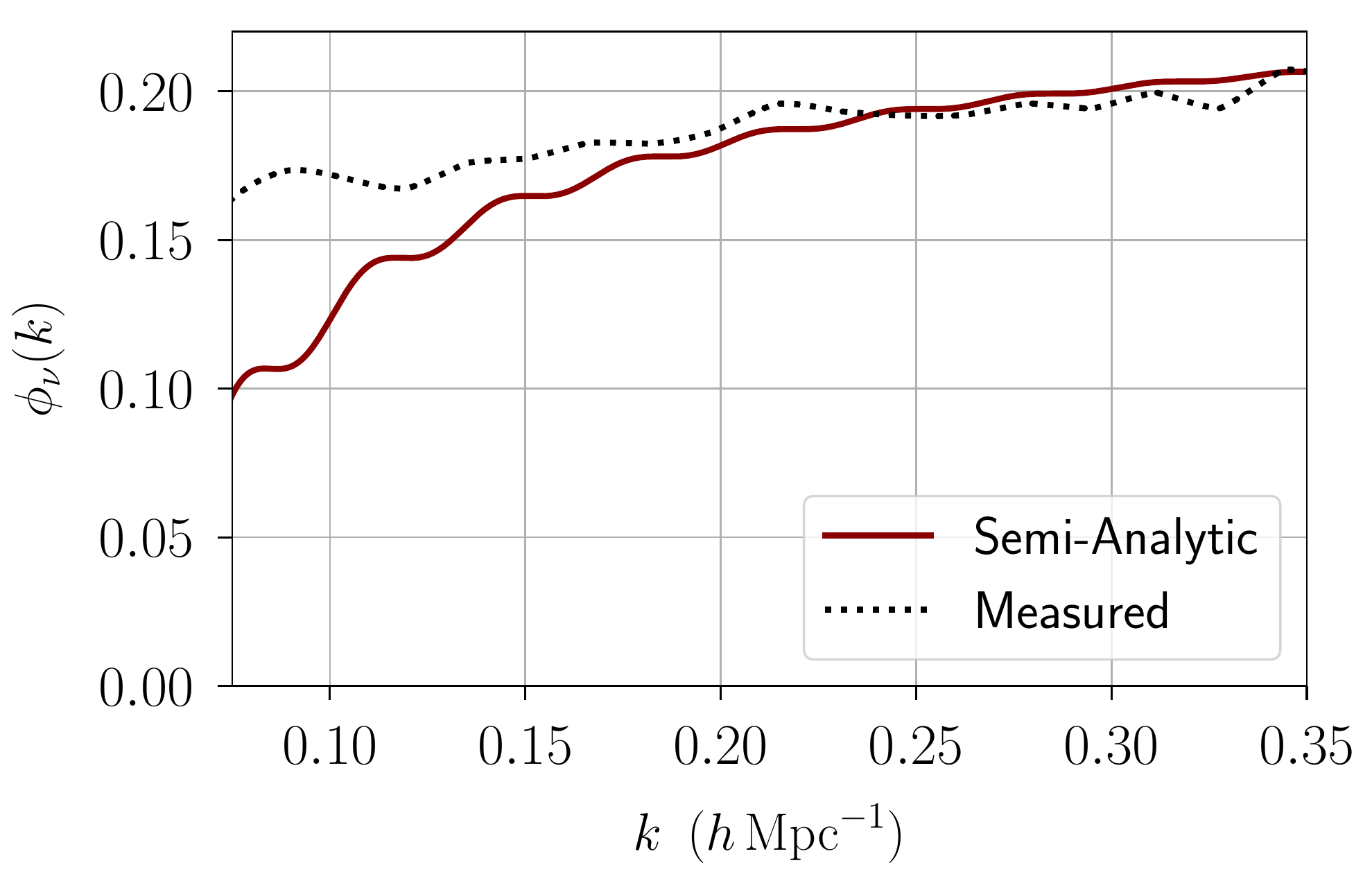}
\caption{Phase shift due to neutrino anisotropic stresses.  The solid curve is the semi-analytic model of the phase shift obtained in (\ref{neutrino phase shift}).  The dotted line is the phase shift measured in~\cite{Baumann:2017gkg} using the output of CLASS for different values of $N_{\rm eff}$, holding the frequency and matter-radiation equality fixed.  The measured value does not uniquely isolate the contribution from anisotropic stress and thus it is possible the mild disagreement at low-$k$ is simply a difference in how $\phi_\nu(k)$ is defined in each case.}
\label{fig:phinu}
\end{figure}  

Equations (\ref{fs for phase}) and (\ref{f nus pieces}) imply the phase shift of $\Theta_0$ due to neutrino anisotropic stresses is
\begin{align}
\label{neutrino phase shift}
    \phi_\nu(k) = {\rm sin}^{-1}(B_{\nu}(k,\eta_*)).
\end{align}
As $k \rightarrow \infty$, the phase shift approaches a constant~\cite{Bashinsky:2003tk,Baumann:2015rya}.  The leading scale dependence is given by the simple power law $1/(kr_s)$, while the $1/(kr_s)^2$ correction adds an oscillatory profile on top of the leading behavior.  This is depicted in Figure \ref{fig:phinu}, which plots (\ref{neutrino phase shift}) against a numerical evaluation of $\phi_\nu$~\cite{Baumann:2017gkg}.  The figure suggests that the BAO phase shift due to neutrino anisotropic is nearly constant for $k \sim 0.3h {\rm Mpc}^{-1}$.

\subsection{Higher-Order in Tight Coupling}
\label{tight coupling}

Up to this point we have worked to leading order in the tight coupling approximation. Terms higher order in $k/\dot{\tau}$ represent the fact that individual photons travel a finite distance between scatterings and undergo a random walk~\cite{Zaldarriaga:1995gi,Hu:1995en}.  The photon diffusion length is given by the average distance of the random walk and sets the scale below which correlations in the photon temperature perturbation are washed out.  Perturbations at distance scales below the diffusion length are suppressed and, as a result, the BAO amplitude is smaller at these scales. 

Higher order terms in the tight coupling approximation then contribute damping terms to (\ref{photon monopole sol}).  The change in the photon oscillations is analogous to what happens when a damping force is added to a harmonic oscillator.  In particular, the amplitude decreases and the phase gets shifted for wavevectors comparable to the inverse diffusion length.  In this section, we follow the discussion of~\cite{Pan:2016zla} to compute the damping factor and phase shift.

During tight coupling, the photon multipole moments and baryon velocity oscillate at the same frequency and can be written
\beq
\Theta_n, v_b \propto e^{i \int^t dt' \omega(t')} \ .
\eeq
To compute the damping factor and phase shift resulting from corrections to tight coupling, the oscillation frequency $\omega(\eta)$ needs to be computed to $O(k/\dot{\tau})^2$. Including the contribution from the photon quadrupole, the equation of motion for $\Theta_1$ (\ref{photon EOM moments multi}) can be written 
\beq
\label{theta1 tight improved theta2}
\Theta_{1}-\frac{v_{b}}{3} =\frac{1}{\dot{\tau}}\left[i \omega \Theta_{1}+k\left(\frac{2}{3} \Theta_{2}-\frac{1}{3} \Theta_{0}\right)\right].
\eeq
Due to the factor of $1/\dot{\tau}$ multiplying the right hand side of (\ref{theta1 tight improved theta2}), the left hand side needs to be computed to $O(k/\dot{\tau})^3$.  The equation of motion for the baryon velocity can be used to express $v_b$ in terms of $\Theta_1$.  Neglecting terms proportional to $\dot a/a$ and $\Psi$ in equation~(\ref{baryons EOM}), the baryon velocity can be written 
\beq
v_b \simeq 3 \Theta_1 \left( 1 - i \frac{\omega R}{\dot \tau} \right)^{-1}.
\eeq
Expanding in $\omega/\dot \tau$, we find
\begin{align}
&\Theta_{1}-\frac{v_{b}}{3} =\Theta_{1}\left[-i \frac{\omega R}{\dot{\tau}}+\left(\frac{\omega R}{\dot{\tau}}\right)^{2}+i\left(\frac{\omega R}{\dot{\tau}}\right)^{3} \right] \ .
\end{align}
The equation for $\Theta_2$ can be obtained by taking the second multipole moment of (\ref{free stream neuts}).  Including the contribution from fluctuations in the photon polarization distribution, this gives
\beq
\label{theta0 theta2 vb tight}
\begin{aligned} 
\Theta_{2} =-\frac{8}{15} \frac{k}{\dot{\tau}} \left(1+\frac{11}{6} \frac{i \omega}{\dot{\tau}}\right)\Theta_{1}.
\end{aligned}
\eeq

Combining $i\omega\Theta_0 = -k\Theta_1$ with equations (\ref{theta1 tight improved theta2}) through (\ref{theta0 theta2 vb tight}) gives the dispersion relation of the photon-baryon system to $O(k/\dot{\tau})^2$:
\begin{align}
\label{frequency equation}
i\omega(1+R)-\frac{ik^2}{3\omega}-\frac{(\omega R)^2 + \frac{16k^2}{45}}{\dot{\tau}}-i\frac{(\omega R)^3+\frac{88}{135}k^2\omega}{\dot{\tau}^2} = 0.
\end{align}
Plugging in the ansatz
\begin{align}
\omega = \omega_0 + i\frac{k}{\dot{\tau}}\gamma + \frac{k^2}{\dot{\tau}^2}\delta \omega_0
\end{align} 
and solving order by order in $k/\dot{\tau}$ gives 
\beq
\begin{aligned} \omega_{0} &=k c_{s},\ \frac{\gamma}{\omega_{0}}=- \frac{\left(c_{s}^{2} R^{2}+\frac{16}{45}\right)}{2 c_{s}\left(1+R\right)},\ \frac{\delta \omega_{0}}{\omega_{0}} = \frac{(\frac{88}{135} + c_s^2 R^3) + (2c_s R^2-\frac{1}{3c_s^2}\frac{\gamma}{\omega_0})\frac{\gamma}{\omega_0}}{2(1+R)}
\end{aligned}
\eeq
where again $c_s^{-1} \equiv \sqrt{3(1+R)}$.  Including corrections to tight coupling, the photon monopole is then
\begin{align}
\label{photon monopole with damping and tight phase shift}
\Theta_0(k,\eta) + \Phi(k,\eta)\simeq - {\rm cos}(kr_s(\eta) + \phi_{\rm dcp}(\eta))e^{-k^2/k^2_D(\eta)}
\end{align}
where the damping factor and phase shift are given by
\beq
\begin{aligned}
\label{tight coupling damp and phase shift}
&k^2/k^2_D(\eta) \equiv k\int_{0}^\eta d\eta'\frac{\gamma}{\dot{\tau}},\ \phi_{\rm dcp}(\eta) \equiv k^2 \int_{ 0}^\eta d\eta' \frac{\delta\omega}{\dot{\tau}^2}.
\end{aligned}
\eeq
Corrections to tight coupling induce a leading order damping factor and a second order phase shift.  Taking the time derivative of (\ref{photon monopole with damping and tight phase shift}) yields an expression for the photon dipole
\begin{align}
\label{photon dipole damped}
\Theta_1(k,\eta) = -c_s\sqrt{1 + \left(\frac{\gamma}{c_s \dot{\tau}}\right)^2} {\rm sin}(kr_s(\eta) + \phi_{\rm dcp}(\eta) + \phi_1(\eta))e^{-k^2/k^2_D(\eta)},
\end{align}
where
\begin{align}
\label{dipole phase shift diffusion}
\phi_1(\eta) \equiv {\rm tan}^{-1}\left(\frac{\gamma}{c_s \dot{\tau}}\right)
\end{align}
is an additional phase shift due to diffusion damping.  This implies the photon monopole and dipole are not $90$ degrees out of phase from one another in the presence of diffusion damping.  Note, the phase shifts $\phi_{\rm dcp}$ and $\phi_1$ become larger for increasing $k$.  Consequently, corrections to tight coupling induce a scale dependent phase shift in the BAO profile which is largest at the upper end of the observable range.

\subsection{Improved Tightly Coupled Photons}
\label{combine improved tight coupling}

Equations (\ref{phase shift grav driving}), (\ref{neutrino phase shift}), and (\ref{tight coupling damp and phase shift}) give the phase shifts to the tightly coupled photons from gravitational driving during matter domination, neutrino anisotropic stresses and higher order corrections to the tight coupling approximation.  Combining terms, we obtain an improved expression for the tightly coupled photon monopole, which is valid until the photons begin to decouple from the baryons,
\begin{align}
\label{Improved photon oscillations in tight couple}
\Theta^{\rm tight}_0(k,\eta)
+\Phi(k,\eta) &= \left[f_\Phi(k,\eta) - \mathcal{A}_{\rm tight}(k){\rm cos}(k r_s(\eta) + \phi_{\rm tight}(k))\right]e^{-k^2/k^2_D(\eta)}
\end{align}
where we have defined
\beq
\begin{aligned}
\label{photon amplitude and phase}
&\mathcal{A}_{\rm tight}(k) \equiv \sqrt{(1 - A_{\rm eq}(k) - A_{\Phi}(k) - A_{\nu}(k))^2 + (B_{\rm eq}(k) +B_{\Phi}(k)+ B_{\nu}(k))^2}\cr
&\phi_{\rm tight}(k,\eta) \equiv {\rm sin}^{-1}(B_{\rm eq}(k) + B_{\Phi}(k)+ B_{\nu}(k)) + \phi_{\rm dec}(k,\eta).
\end{aligned}
\eeq
$A_{\rm eq}$ and $B_{\rm eq}$ are corrections due to the evolution during radiation domination and are derived in Appendix \ref{rad matt matching app}.  The photon dipole can be obtained from (\ref{Improved photon oscillations in tight couple}) using the relation $\Theta_1 = -\dot{\Theta}_0/k$. 
Equation (\ref{Improved photon oscillations in tight couple}) sets the initial conditions for the evolution of the photons and the force they exert on the baryons during decoupling, which is the subject of the next section.

\section{Corrections Near Decoupling}
\label{sec:decouple}

In section \ref{baseline sec}, we made a number of simplifying assumptions regarding the dynamics near decoupling.  In particular, we evaluated (\ref{v_b full 2}) using the tightly coupled solution to $\Theta_1$, ignored the skewness of $g_b(\eta)$ by approximating it as a Gaussian, and neglected the term proportional to $\delta_b(\eta_g)$ in (\ref{matter growing mode coeff}).  In this section, we compute the leading corrections to these approximations and combine them with the results of the previous section to derive an improved expression for the BAO.

\subsection{Transients}
\label{transient section}

Near $\eta \sim \bar{\eta} \equiv 250 {\rm Mpc}^{-1}$, the photon-baryon scattering rate has decreased and $\dot{\tau} \sim k$ for the observable BAO modes.  Photons can no longer be treated as tightly coupled to the baryons and the photon oscillation frequency deviates from $kc_s$.  Even though the photons and baryons are not tightly coupled, transient photon-baryon interactions continue to influence the photon evolution until the photons fully decouple from the baryons. 
\begin{figure}
\centering
\includegraphics[width=5in]{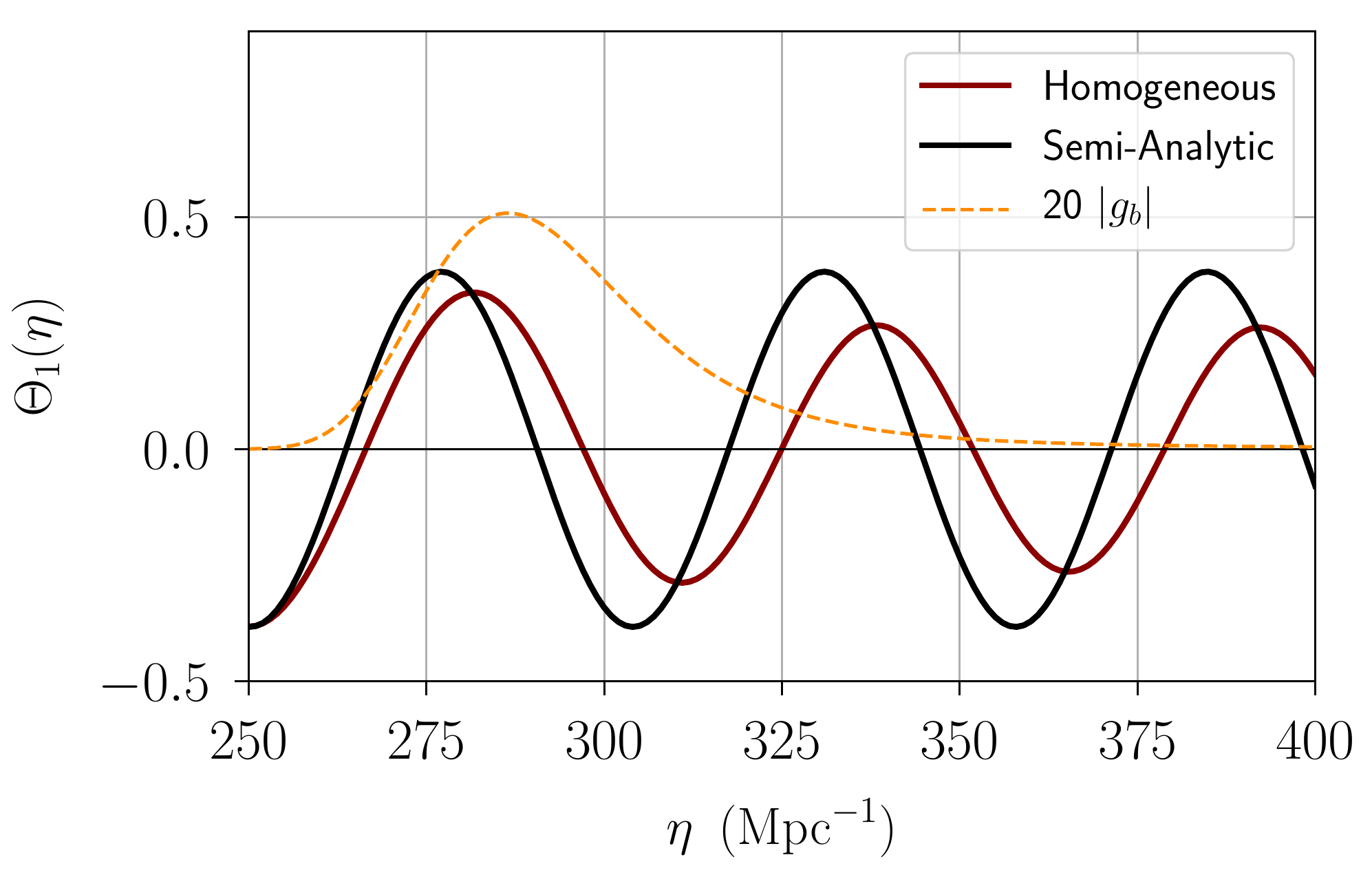}
\caption{ The effects of transient photon-baryon scatterings on $\Theta_1$ for $k = 0.3h {\rm Mpc}^{-1}$.  The red curve includes transients and depicts a numerical evaluation of (\ref{trans phot eom2}).  The black curve ignores the effects of transients and depicts the homogeneous part of (\ref{Theta1 improved2}).    The orange dotted curve plots the (rescaled) baryon visibility function $g_b(\eta)$.}
\label{fig:transientplot}
\end{figure}
Consider the terms in equations (\ref{baryons EOM}) and (\ref{photon EOM moments multi}) that describe photon-baryon interactions:
\begin{align}
\label{trans phot eom2}
&\ddot{\Theta}_0 + \frac{k^2}{3}\Theta_0 =  \frac{kR}{3}\dot{v}_b,\ \quad\ \dot{v}_b =  \dot{\tau}_b(v_b - 3\Theta_1).
\end{align}
Since the photons and baryons are not tightly coupled $v_b \neq 3\Theta_1$.  The relative photon-baryon motion induces a drag force on the photon oscillations, which alters their frequency and gives rise to a phase shift.  

Ignoring free-streaming effects for the moment, the photon dipole for $\eta > \bar{\eta}$ can be expressed as
\begin{align}
\label{Theta1 improved2}
\Theta_1(\eta) &= (Ce^{i\omega(\eta-\bar{\eta})}+c.c.) + \Theta^{\rm trans}_1(\eta)
\end{align}
where $\omega \equiv k/\sqrt{3}$.  The first term is the homogeneous solution of (\ref{trans phot eom2}) and $\Theta^{\rm trans}_1(\eta)$ parametrizes the correction from transients.  
The constant $C$ encodes the phase and amplitude of the photons as they transition from the tightly coupled regime to the transient one, and is fixed by matching (\ref{Theta1 improved2}) to (\ref{Improved photon oscillations in tight couple}) at $\bar{\eta}$:
\begin{align}
\label{C match trans tight2}
C = -\frac{i}{2\sqrt{3}}(\Theta^{\rm tight}_0(\bar{\eta}) - \frac{i}{\omega}\dot{\Theta}^{\rm tight}_0(\bar{\eta})) \equiv  i|\bar{C}|e^{i(kr_s(\bar{\eta}) + \phi_{\rm tight}(\bar{\eta}) + \phi_{\rm match} + \phi_1(\bar{\eta}))}e^{-k^2/k_D^2(\bar{\eta})}
\end{align}
where $\phi_{\rm match}$ is defined as the phase shift due to the matching and $|\bar{C}|$ is a real number. 

The contribution from transients $\Theta^{\rm trans}_1(\eta)$ is determined by solving (\ref{trans phot eom2}).  While these equations are difficult to solve analytically, the qualitative effects of transients can readily be understood by studying the numerical solution.  Figure \ref{fig:transientplot} plots a numerical evaluation of (\ref{trans phot eom2}) against the homogeneous piece of (\ref{Theta1 improved2}).  We see that the effect of transients is to increase the photon oscillation period.  The curve including transients lags behind the homogeneous one, and accumulates a phase shift by the time transients have dissipated.  As seen in figure \ref{fig:transientplot}, the phase difference is appreciable by the peak of the baryon visibility function.  As a result, transients induce a phase shift in the BAO profile.  

It turns out that the lifetime of transients $\ddot{\tau}/\dot{\tau}$ is comparable to the oscillation frequencies of modes in the middle of the observable BAO range, $k\sim 0.2 \, h \, {\rm Mpc}^{-1}$.  Modes whose oscillation periods are longer than the transient lifetime should be less sensitive to the effects of transients.  We then expect the size of the BAO phase shift due to transients to be an increasing function of $k$.

\subsection{Free-Streaming}
\label{sec:quadrupole}
As discussed in the previous section, photons are no longer tightly coupled to the baryons after $\bar{\eta}$.  Since the photon-baryon scattering rate has decreased, photons begin to free-stream and higher photon multipole moments cannot be neglected.  One way to account for free-streaming is by integrating (\ref{free stream neuts}) to obtain an expression for the full photon temperature fluctuation $\Theta$, and then project out $\Theta_1$ as is done in the analysis of the CMB.  However, this description hides the fact that free-streaming only induces a perturbative correction to the BAO profile.  The BAO is most sensitive to the photon behavior during the era of decoupling, before photons fully enter the free-streaming regime.  As a result, contributions from free streaming are better described as a perturbative correction to the (truncated) fluid description where $\Theta_2=0$.

The fact that free-streaming effects only provide a small correction to the BAO can be understood by noting that the total comoving distance a free-streaming photon travels during decoupling is of order $\sigma$, the width of the baryon visibility function.  This turns out to be less than the comoving wavelengths of observable BAO modes $\lambda \sim 2\pi/k$.  Since the photons were originally tightly coupled, individual photons do not travel far during the era of decoupling relative to the BAO length scales.  The truncated fluid description is then sufficient for the purposes of determining the BAO, and the effects of free-streaming can be treated as a perturbation.

To derive the corrections from free streaming, we include the term proportional to $\Theta_2$ in equation (\ref{eq:box_theta}),
\begin{align}
\label{monopole equation with quad and trans}
\ddot{\Theta}_0 + \frac{k^2}{3}\Theta_0 = \frac{kR}{3}\dot{v}_b + \frac{2k^2}{3}\Theta_2 \ .
\end{align}
The solution to (\ref{monopole equation with quad and trans}) can be separated into contributions coming from the homogeneous evolution, transients, and free streaming.  Using $\Theta_1 = -\dot{\Theta}_0/k$, we find
\begin{align}
\label{dipole including hom trans free}
\Theta_1(\eta) = (Ce^{i\omega(\eta-  \bar{\eta})}+c.c.) + \Theta^{\rm trans}_1(\eta) + \Theta^{\rm FS}_1(\eta)
\end{align}
The behavior of $\Theta^{\rm trans}_1(\eta)$ is most sensitive to the dynamics near $\bar{\eta}$, when the transient photon-baryon interactions are largest.  Residual photon-baryon scatterings suppress free-streaming around this time.  To a good approximation then $\Theta^{\rm trans}_1(\eta)$ can be computed using (\ref{trans phot eom2}).  

The contribution from free-streaming can be expressed as
\begin{align}
\label{monopole including quadrupole source}
\Theta^{\rm FS}_1(\eta) \equiv -\frac{k}{3}\int_{\bar{\eta}}^\eta d\eta' e^{i\omega(\eta-\eta')}\Theta_2(\eta') + c.c. 
\end{align}
We can obtain an expression for $\Theta_2$ by first integrating (\ref{free stream neuts}) for the full photon perturbation
and projecting out the quadrupole component.  Neglecting contributions from the gravitational potentials, we find 
\begin{align}
\label{full photon quadrupole free}
\Theta_2(\eta) &\simeq e^{\tau(\eta) - \tau(\bar{\eta})}(\Theta_0(\bar{\eta})j_2(k(\eta-\bar{\eta}))+3\Theta_1(\bar{\eta})j'_2(k(\eta-\bar{\eta})))\cr
&\ \ \ \ -\int_{\bar{\eta}}^\eta d\eta' e^{\tau(\eta) - \tau(\eta')}\dot{\tau}(\eta')(\Theta_0(\eta')j_2(k(\eta - \eta')) + v_b(\eta')j_2'(k(\eta - \eta'))).  
\end{align}
Due to the factor of $\dot{\tau}(\eta')$, the majority of the integrand's support comes from the period of time when transient photon-baryon interactions are largest.  The integral can then be evaluated using the solutions to (\ref{trans phot eom2}).  

To obtain the free-streaming correction to $\Theta_1$, we evaluate (\ref{full photon quadrupole free}) numerically and plug the result into (\ref{monopole including quadrupole source}).  Similar to what we found in section \ref{tight coupling}, free-streaming dampens the photon driving force exerted on the baryons and decreases the BAO amplitude.  In addition, it induces a scale-dependent phase shift in the BAO profile which becomes larger for modes comparable to the free-streaming length during decoupling $k \sim 2\pi/\sigma$.  The correction to the BAO profile due to free-streaming effects should then be largest for modes at the upper end of the observable BAO range.

\subsection{Baryon Visibility Function and Baryon Velocity}
\label{baryon velocity improved section}
\begin{figure}
\centering
\includegraphics[width=5in]{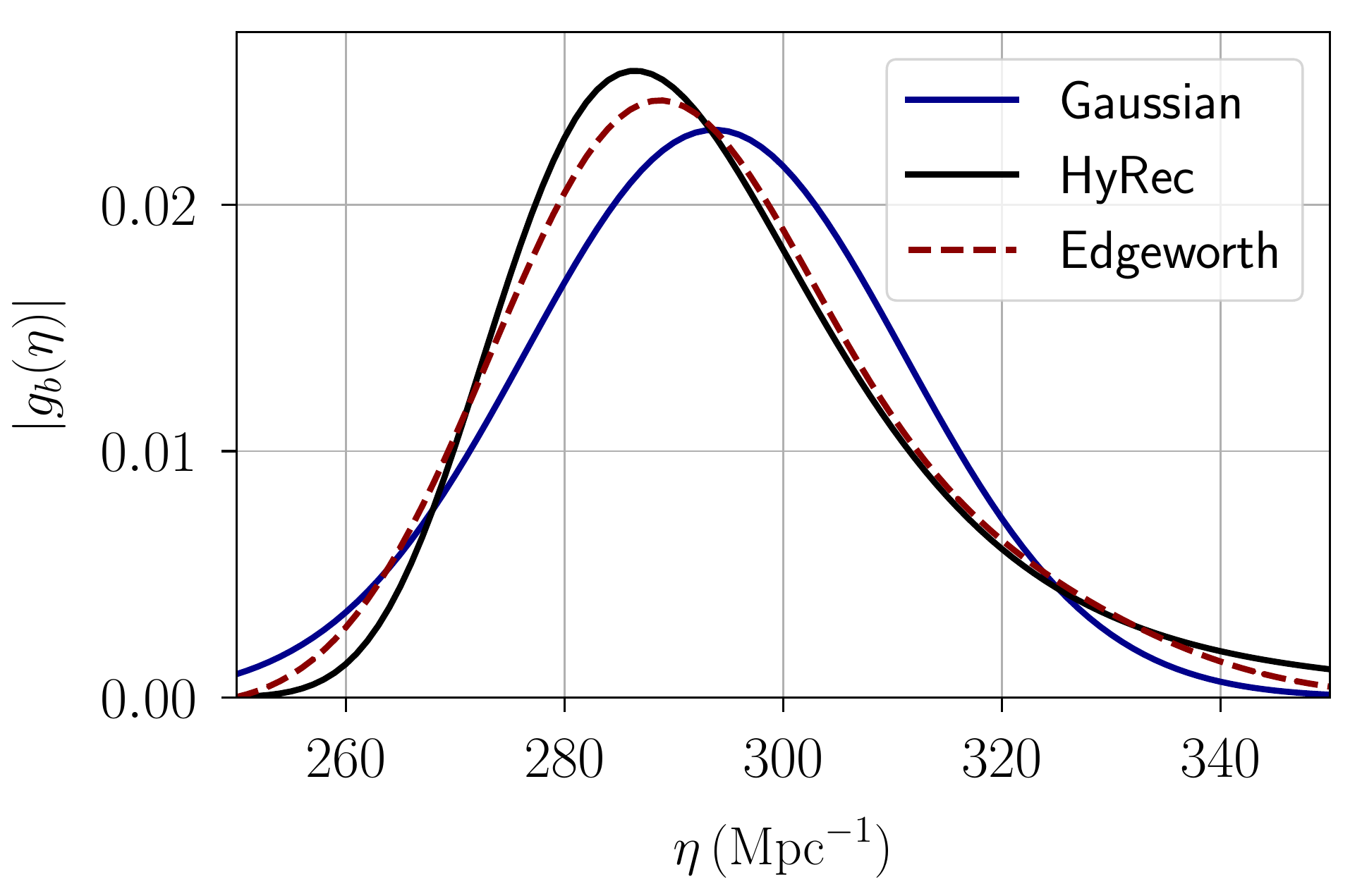}
\caption{Baryon visibility function $g_b(\eta)$.  The black curve is obtained numerically from HyRec~\cite{AliHaimoud:2010dx}, while the blue and red curves depict (\ref{edgeworth}) with and without the skewness parameter $k_3$.}
\label{fig:gbplotcompare}
\end{figure}

In section \ref{baseline sec}, the baryon visibility function $g_b(\eta)$ was approximated by a Gaussian to obtain $v_b(\eta_g)$.  However, this fit does not capture the asymmetric shape of $g_b$, which itself induces a phase shift in the BAO.  The Gaussian approximation of $g_b$ can be improved using a Gaussian Edgeworth expansion.  To leading order
\begin{align}
\label{edgeworth}
g_b(\eta) \simeq -\frac{1}{\sigma\sqrt{\pi}} e^{-\frac{(\eta - \eta_*)^2}{\sigma^2}}(1+ k_3 H_3\left(\frac{\eta-\eta_*}{\sigma}\right)),\ k_3 = -\frac{\sigma\sqrt{\pi}\int d\eta g_b(\eta)H_3(\frac{\eta-\eta_*}{\sigma})}{\int d\eta e^{-\frac{(\eta-\eta_*)^2}{\sigma^2}}H_3(\frac{\eta-\eta_*}{\sigma})^2}
\end{align}
where $H_3(x)$ denotes the third Hermite polynomial and $k_3$ parametrizes the skewness of $g_b$.  Figure \ref{fig:gbplotcompare} plots $g_b$ obtained numerically against (\ref{edgeworth}) and the Gaussian approximation.  Clearly (\ref{edgeworth}) gives a more accurate fit than the Gaussian approximation.  In particular, it does a better job of approximating the peak location and fitting the slowly decaying tail of $g_b$.

We are now in a position to derive an improved formula for $v_b(k,\eta_g)$.  By combining the full expression for the photon dipole valid during decoupling (\ref{dipole including hom trans free}) with (\ref{v_b full 2}), we find
\begin{align}
\label{full vb baryon vis integrate}
v_b(k,\eta_g) = -3\frac{a(\eta_*)}{a(\eta_g)}\int_{-\infty}^{\infty} d\eta g_b(\eta)\left[(Ce^{i\omega(\eta-\bar{\eta})}+c.c.) + \Theta^{\rm trans}_1(\eta) + \Theta^{\rm FS}_1(\eta) \right] + I_\Phi(k,\eta_g)
\end{align}
The contributions from $\Theta^{\rm trans}_1(\eta)$ and  $\Theta^{\rm FS}_1(\eta)$ can be evaluated numerically.  The piece resulting from the homogeneous evolution of $\Theta_1$ can be obtained analytically using (\ref{edgeworth}).  It is straightforward to evaluate the time integral and find
\begin{align}
\label{edgeworth exponential integral}
\int_{-\infty}^{\infty}d\eta' g_b(\eta')e^{i\omega(\eta'-\bar{\eta})} \simeq -\sqrt{1 + k^2_3 (\sigma\omega)^6}e^{-\frac{\omega^2\sigma^2}{4}}e^{i\omega(\eta_* - \bar{\eta}) + i\phi_{\rm skew}}
\end{align}
where $\phi_{\rm skew} \equiv -{\rm tan}^{-1}(k_3 (\sigma \omega)^3)$ represents the phase shift due to the skewness of $g_b$.

Plugging (\ref{C match trans tight2}) and (\ref{edgeworth exponential integral}) into (\ref{full vb baryon vis integrate}) gives the baryon velocity at $\eta_g$.  We find
\begin{align}
\label{vb at grow improved}
v_b(k,\eta_g) =  - \mathcal{A}_{v_b}{\rm sin}(kr_s(\eta_*) + \phi_{v_b}) + v^{\rm trans}_b(k,\eta_g) +  v^{\rm FS}_b(k,\eta_g) + I_{\Phi}(k,\eta_g)
\end{align}
where $v^{\rm trans}_b$ and $v^{\rm FS}_b$ parametrize the corrections due to transients and free-streaming.  The amplitude and phase of the piece due to the homogeneous evolution of $\Theta_1$ are
\begin{align}
\mathcal{A}_{v_b} &\equiv 6|\bar{C}| \frac{a(\eta_*)}{a(\eta_g)}\sqrt{1 + k^2_3 (\sigma\omega)^6} e^{-k^2/k^{2}_D(\bar{\eta})}e^{-\frac{\omega^2 \sigma^2}{4}}\cr
\phi_{v_b} &\equiv k(r_s(\bar{\eta}) - r_s(\eta_*) - \frac{\bar{\eta} - \eta_*}{\sqrt{3}})  + \phi_{\rm tight}(\bar{\eta}) + \phi_{\rm match} + \phi_1(\bar{\eta}) + \phi_{\rm skew}.
\end{align}
The first contribution to $\phi_{v_b}$ is due to the fact the photons no longer propagate at the speed $c_s$ after $\bar{\eta}$.  The remaining contributions come from corrections to the tightly coupled photons (\ref{photon amplitude and phase}), matching at $\bar{\eta}$ and the skewness of $g_b$.  The factors of $v^{\rm trans}_b(k,\eta_g)$ and $v^{\rm FS}_b(k,\eta_g)$ in (\ref{vb at grow improved}) encode phase shifts due to transients and free-streaming.  The amplitude of $v_b$ is suppressed at large $k$ by diffusion damping.  In addition, free streaming during decoupling also decreases the amplitude of $v_b(k,\eta_g)$.

\subsection{Baryon Overdensity}\label{sec:deltab}

As described in section \ref{baseline sec}, the baryon velocity at $\eta_g$  gives the leading contribution to the BAO profile. This is due to the velocity overshoot effect, i.e. the fact that the term proportional to $\delta_b$ in the expression for $C_{\rm BAO}$ (\ref{matter growing mode coeff}) is suppressed relative to the term proportional to $v_b$ by a factor of $aH/k$.  However, in the time between $\bar{\eta}$ and $\eta_g$, the baryons have begun to experience gravitational growth and the magnitude of $\delta_b$ has increased several times larger than the magnitude of $v_b$ by $\eta_g$.  This partially compensates for the small factor multiplying $\delta_b$, which means the contribution from $\delta_b$ in (\ref{matter growing mode coeff}) gives a leading correction to the BAO profile.  To derive $\delta_b(k,\eta_g)$, we integrate $\dot{\delta}_b = -k v_b$ from $\bar{\eta}$ to $\eta_g$ and find
\beq
\begin{aligned}
\label{delta g at g}
\delta_b(k,\eta_g) &\simeq 3\Theta_0(k,\bar{\eta}) + 3k\int_{\bar{\eta}}^{\eta_g}d\eta \frac{e^{\tau_b(\eta)}}{a(\eta)}\int_{-\infty}^{\eta} d\eta' a(\eta')g_b(\eta') \Theta_1(\eta')- k\int_{\bar{\eta}}^{\eta_g}d\eta' I_\Phi(k,\eta')
\end{aligned}
\eeq
where we used $\delta_b(\bar{\eta}) \simeq 3\Theta_0(\bar{\eta})$.  It is straightforward to evaluate (\ref{delta g at g}) numerically.

\section{BAO}\label{sec:BAO}
Plugging our expressions for the baryon velocity (\ref{vb at grow improved}) and overdensity (\ref{delta g at g}) at $\eta_g$ into (\ref{matter growing mode coeff}) yields an improved expression for the BAO
\beq
\tcboxmath{\begin{aligned}
\label{Full BAO analytic}
C_{\rm BAO}(k) &= \frac{2k}{5\dot{y}(\eta_g)}\frac{\Omega_b}{\Omega_m}\left(\mathcal{A}_{v_b}(k){\rm sin}(kr_s(\eta_*) + \phi_{v_b}(k))- v^{\rm trans}_b(k,\eta_g) \right.\cr
&\left.\ \ \ \ \ \ \ \ \ \ \ \ \ \ \ \ \ \ \ \ \ \ \ \ \ \ - v^{\rm FS}_b(k,\eta_g)+\frac{3aH(\eta_g)}{2k}\delta_b(k,\eta_g)-I_\Phi(k,\eta_g)\right).
\end{aligned}}
\eeq
The phase shift $\phi_{v_b}$ is due to corrections to the tightly coupled photons (described in section~\ref{sec:tight}) and the skewness of the baryon visibility function (section~\ref{baryon velocity improved section}).  The contributions from transient photon-baryon interactions and free-streaming (sections~\ref{transient section} and~\ref{sec:quadrupole}) are parametrized by $v^{\rm trans}_b$ and $v^{\rm FS}_b$.  Finally, the contributions from the baryon overdensity $\delta_b$ and gravitational potential $I_\Phi$ are described in sections~\ref{sec:deltab} and~\ref{baseline sec}.

\begin{figure}
\centering
\includegraphics[width=5in]{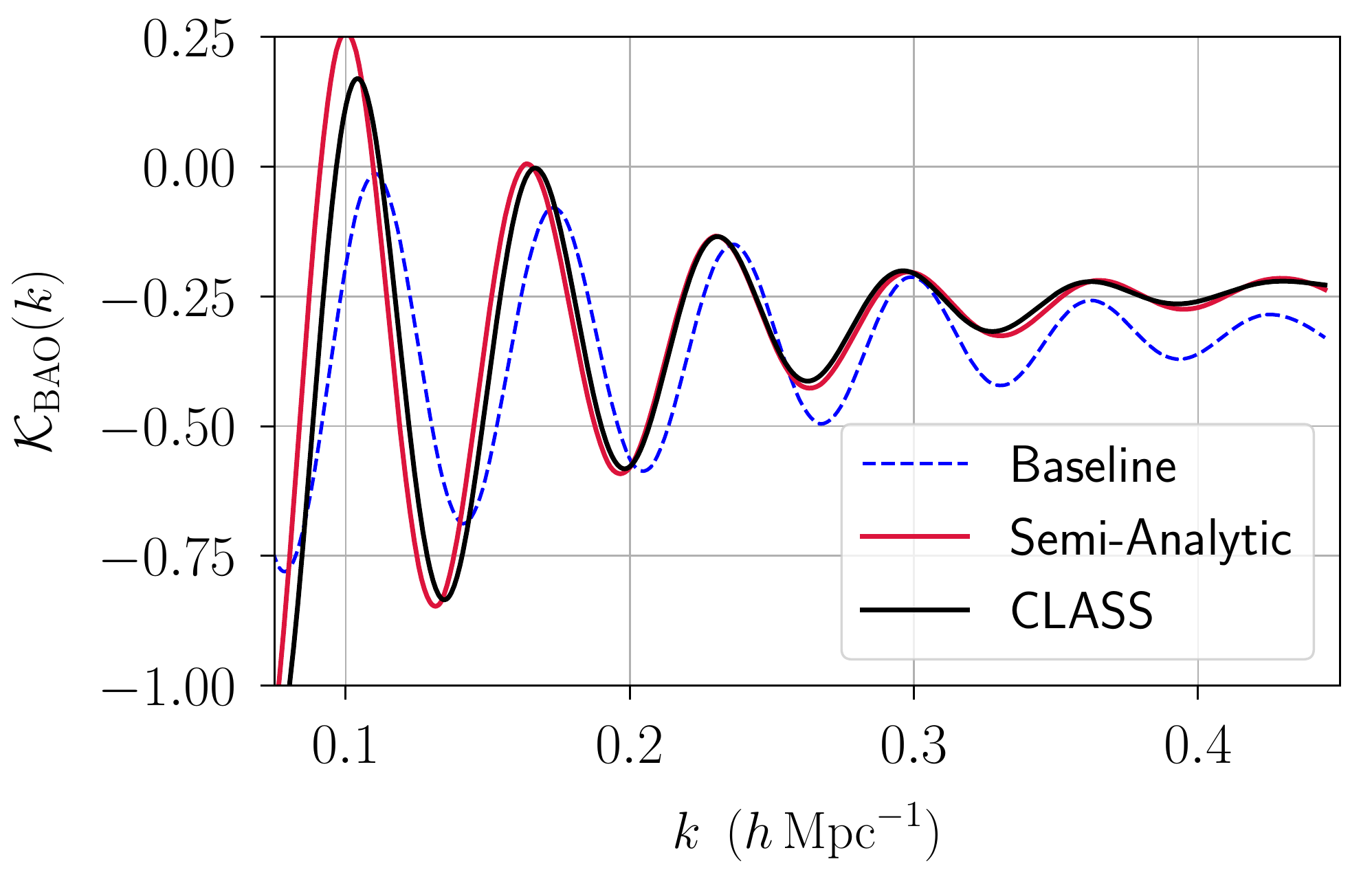}
\caption{Plot of the baryon growing mode coefficient ${\cal K}_{\rm BAO} = C_{\rm BAO}/(\frac{2k}{5\dot{y}(\eta_g)}\frac{\Omega_b}{\Omega_m})$ given by equation (\ref{Full BAO analytic}) (red) versus the numerical result (black) obtained with CLASS~\cite{Blas:2011rf}.  The blue curve is the baseline curve describe in equation~(\ref{zeroth order BAO}), after modifying its amplitude and offset from zero to agree with the amplitude of the third peak and trough obtained with CLASS.}
\label{fig:BAOfull}
\end{figure}

Figure \ref{fig:BAOfull} plots $C_{\rm BAO}$ obtained numerically from CLASS~\cite{Blas:2011rf} against the baseline model (\ref{zeroth order BAO}) and our improved semi-analytic model (\ref{Full BAO analytic}).  The breakdown in the agreement between the semi-analytic model and the output from CLASS at lower $k$ is expected.  Our expressions for the contributions from gravitational driving during matter domination and the phase shift due to neutrino anisotropic stress (sections \ref{gravity driving during matter dom} and \ref{Anisotropic stress section}) rely on $kr_s(\eta_{\rm eq}) \gg 1$ and $kr_s(\eta_*) \gg 1$, which is more accurate for larger $k$.

Table~\ref{deltaktab} lists the changes to the first four BAO peak locations in Fourier space due to (from left to right) gravitational driving during matter domination, neutrino anisotropic stresses, higher order corrections to tight coupling, transient photon-baryon interactions, free-streaming of photons, the skewness of the baryon visibility function and the baryon overdensity at $\eta_g$.  The variations are determined by individually removing each contribution from (\ref{Full BAO analytic}) and subtracting the new BAO peak locations from the old ones.  For comparison, the last column gives the shifts due to a 1$\%$ shift in the sound horizon at $\eta_*$.

As expected, the higher BAO peaks are more sensitive to corrections to tight coupling, transients and free-streaming.  The impact of gravitational driving during matter domination is largest in the first peak location.  The skewness of $g_b$ induces the smallest shifts to the peak locations.  The shifts due to the baryon overdensity at $\eta_g$ are of similar size as $\delta k_\nu$, which is nearly scale-invariant. 

The first three entries in Table~\ref{deltaktab} (left) directly alter the photon distribution around the time of recombination.  As a result, these shifts similarly alter the acoustic peaks of the CMB.   On average, the neutrino phase is the largest of these effects.  At first sight, the weak $k$-dependence of $\delta k_\nu$ might appear to contradict the $k$-dependence observed in Figure~\ref{neutrino phase shift}.  However,  because the amplitude of the BAO is also $k$-dependent, the locations of the peaks and the change to the phase are not equivalent.  The next four effects in the table (middle) are unique to the BAO  and are potentially larger than the neutrino induced phase.  These effects are uniquely sensitive to physics during the decoupling of photons and baryons.  It would be interesting to explore how these effects might be sensitive to beyond $\Lambda {\rm CDM}$ physics.

The total phase shifts are compared to a one-percent shift in $r_{\rm drift}$ (right columns).  Current observations constrain $D_M(z) / r_{\rm drift}$~\cite{Alam:2020sor} at the percent level and thus are likely already sensitive to these shifts.  Next generation surveys such as DESI~\cite{Aghamousa:2016zmz} and Euclid~\cite{Amendola:2016saw} are expected to reach sub-percent level of sensitivity over a larger number of redshift bins, likely translating to a factor of several in  improvement in overall sensitivity~\cite{Font-Ribera:2013rwa}.  These expectations are consistent with forecasts for the measurement of the neutrino phase shift in  current and future surveys~\cite{Baumann:2017gkg}.

\begin{table}
\centering
\begin{tabular}{|c| S S S| S S S S|S||S|}
\hline
\hline
peak no. & $\delta k_{\Phi}$ & $\delta k_{\nu}$ & $\delta k_{\rm tight}$ & $\delta k_{\rm trans}$ & $\delta k_{\rm free}$ & $\delta k_{\rm skew}$ & $\delta k_{\delta_b}$ & $\delta k_{\rm tot}$ & $\delta k_{\text{1\%}}$ \\
\hline
1st & 3 & 3 & 1 & -3 & 1 & -1 & -4 & 0 & 1\\
2nd & 0 & 3 & 1 & -3 & 0 & -1 & -3 & -3 & 2\\
3rd & 0 & 3 & 2 & -6 & -4 & 1 & -3 & -7 & 3\\
4th & 1 & 3 & 4 & -8 & -8 & 0 & -3 & -11 & 3\\ 
\hline
\end{tabular}
\caption{BAO peak shifts resulting from corrections to tightly coupled photons and decoupling.  Shifts are in units of $10^{-3}h{\rm Mpc}^{-1}$.  For comparison, the last column gives the shifts in the peak locations due to a 1$\%$ shift in the sound horizon at $\eta_*$.}
\label{deltaktab}
\end{table}

\section{Conclusions and Outlook}\label{sec:conclusions}

In this paper, we presented a detailed analysis of the different physical effects that give rise to the BAO phase in Fourier space, and derived a semi-analytic model for the BAO.  This model was used to determine the variations of the BAO peak locations due to seven corrections to the tightly coupled photons and the physics during decoupling.  The effects from gravitational driving during matter domination, neutrino anisotropic stresses and higher order corrections to the tight coupling approximation are common to the peak locations of both the CMB and the BAO.  The remaining corrections are unique to the physics of baryon decoupling and could, in principle, be altered without impacting the CMB directly.  Some of these shifts are comparable in size to the phase shift due to neutrino anisotropic stress or a one-percent shift in the BAO scale, and thus should be large enough to be measured in current and future data.

Our broader motivation for this work is to understand how physics beyond $\Lambda$CDM can alter the BAO non-trivially.  This is relevant to the use of the BAO as a standard ruler, since measurements of the BAO phase and peak are degenerate~\cite{Baumann:2017gkg,Bernal:2020vbb}.  Our analytic understanding of these phase corrections gives us a window into the types of models that might non-trivially impact the BAO in a way that is not captured by standard BAO analyses.  The importance of this is highlighted by the fact that BAO measurements will improve~\cite{Font-Ribera:2013rwa} to the point where these phases could be measured with high significance~\cite{Baumann:2017gkg}.  It would be particularly interesting to investigate whether a direct measurement of the phase of the BAO could be provide new insights into the era of baryon decoupling and any possible new physics that could manisfest itself at that time.  

\paragraph{Acknowledgements}

We are grateful to Daniel Baumann, Raphael Flauger, Yi Guo, Lloyd Knox, Zhen Pan, Benjamin Wallisch, Matias Zaldarriaga for helpful discussions. We also thank Benjamin Wallisch for providing the numerical template for the neutrino phase shift. D.\,G.~is supported by the US~Department of Energy under grant no.~DE-SC0019035. A.\,K.\,R.~is supported by the US~Department of Energy under grant no.~DE-SC0009919.

\newpage
\appendix
\section{Matching at Radiation-Matter Equality}
\label{rad matt matching app}
In this appendix, we compute the factors of $A_{\rm eq}$ and $B_{\rm eq}$ appearing in (\ref{Improved photon oscillations in tight couple}) which represent the contributions to the tightly coupled photons due to the evolution during radiation domination.  Ignoring the corrections discussed in section \ref{sec:tight}, the expression for the tightly coupled photons during matter domination is the homogeneous solution of (\ref{photon monopole sol}) and can be written
\begin{align}
\label{some label for photons matter dom}
\Theta_0(k,\eta)  + \Phi(k,\eta)  = -(1-A_{\rm eq}(k)){\rm cos}(kr_s(\eta)) + B_{\rm eq}(k){\rm sin}(kr_s(\eta)).
\end{align}
The coefficients $A_{\rm eq}$ and $B_{\rm eq}$ can be fixed by matching (\ref{some label for photons matter dom}) to the expression for $\Theta_0$ valid during radiation domination.  Neglecting contributions from matter for the moment, the photon monopole during radiation domination takes the well-known form
\begin{align}
\label{rad dom theta0}
\Theta^{\rm rad}_0(k,\eta) = \frac{2((k r_s)^2-1){\rm sin}(k r_s)- k r_s((k r_s)^2-2){\rm cos}(k r_s)}{ (k r_s)^3}.
\end{align}
By matching the functional values and derivatives of (\ref{some label for photons matter dom}) with (\ref{rad dom theta0}) at $\eta_{\rm eq}$, we find
\beq
\begin{aligned}
\label{matching coeffs rad}
&A_{\rm eq}(k)=\frac{k r_s(2{\rm sin}(2 k r_s) - k r_s({\rm cos}(2 k r_s)-3))-6{\rm sin}^2(k r_s)}{(kr_s)^4}+ \Phi {\rm cos}(k r_s)\cr
&B_{\rm eq}(k)= \frac{(3-(k r_s)^2){\rm sin}(2 k r_s) - 2 k r_s(-(k r_s)^2 + {\rm cos}(2 k r_s)+2)}{(k r_s)^4} + \Phi {\rm sin}(k r_s)
\end{aligned}
\eeq
where all time dependent functions in (\ref{matching coeffs rad}) are evaluated at $\eta_{\rm eq}$ and $\Phi$ is given by (\ref{matt dom potential 1}).
The phase shift to $\Theta_0$ due to the photon evolution during radiation domination is then approximately
\begin{align}
\label{phi rad dom shift}
\phi_{\rm eq}(k) = {\rm sin}^{-1}(B_{\rm eq}).
\end{align}

Equation (\ref{rad dom theta0}) is only valid if radiation is the only gravitational source during radiation domination.  In reality, by $\eta_{\rm eq}$ gravitational collapse has made $\delta$ much larger than $\Theta_0$ for the relevant scales.  This means that near $\eta_{\rm eq}$, the scale factor and gravitational potentials deviate from their pure radiation domination expressions.  While (\ref{matching coeffs rad}) gives the leading scale dependence of $A_{\rm eq}$ and $B_{\rm eq}$, the gravitational effects of matter give important corrections.  These corrections are difficult to compute analytically, however, it is straightforward to incorporate them by solving (\ref{Potential eqns}), (\ref{matter eqns}) and (\ref{photon EOM moments multi}) numerically through radiation domination.  The results of this are used to evaluate $A_{\rm eq}$ and $B_{\rm eq}$ in (\ref{Improved photon oscillations in tight couple}) and (\ref{photon amplitude and phase}).

\section{Initial Conditions}
\label{initial conditions}
In this appendix, we derive the initial conditions for the gravitational potentials including contributions from anisotropic stress using the method described in \cite{Ma:1995ey}.  Assuming the initial conditions are adiabatic, then at early times the cosmological perturbations can be approximated by the following series solutions,
\begin{align}
\label{series sols post}
&\Phi = \Phi_0 + \Phi_2 (k\eta)^2,\ \Psi = \Psi_0 + \Psi_2 (k\eta)^2, \Theta_0 = \Theta_{00} + \Theta_{01}(k\eta) + \Theta_{02} (k\eta)^2,\cr
&N_0 = N_{00} + N_{01} (k\eta) + N_{02}(k\eta)^2,\ N_1 = N_{11}(k\eta) + N_{12} (k\eta)^2,\ \Theta_1 = \Theta_{11}(k\eta) + \Theta_{12}(k\eta)^2,\cr
&N_2 = N_{22}(k\eta)^2,\ v_b = v_{b1}(k\eta) + v_{b2}(k\eta)^2.
\end{align}
The series coefficients can be obtained by inserting (\ref{series sols post}) into $\dot{\mathcal{N}}_2 = \frac{2}{5}k\mathcal{N}_1$ and the equations of motion for the metric perturbations and cosmological components given in section \ref{cosmological perturbations equation}. Matching order by order in $\eta$, we find
\begin{align}
\label{early time exps}
&\Phi_0 = -(1 + \frac{2}{5}\epsilon_\nu)\Psi_0,\ \frac{v_b}{3}= \Theta_1 = {\mathcal N}_1 = \frac{k \eta}{6}\Psi_0.
\end{align}
It is useful to write the initial conditions for the gravitational potentials in terms of the conserved scalar curvature perturbation $\zeta$.  Using $i k_i \delta T^{0}_i = 4 a k (\rho_\gamma\Theta_1 + \rho_\nu {\mathcal N}_1)$ and (\ref{early time exps}), we find
\begin{align}
\label{scalar curvature pert}
\zeta \equiv \frac{-ik_i\delta T^{0}_i H}{k^2(\rho + P)} + \Phi = -\frac{3}{2}(1 + \frac{4\epsilon_\nu}{15})\Psi_{0}.
\end{align}
The initial conditions for the gravitational potentials are then
\begin{align}
\Phi = \frac{2}{3}(1+\frac{2\epsilon_\nu}{15} )\zeta,\ \Psi = -\frac{2}{3}(1-\frac{4\epsilon_\nu}{15})\zeta.
\end{align}

\clearpage
\phantomsection
\addcontentsline{toc}{section}{References}
\small
\bibliographystyle{utphys}
\bibliography{Nu_LSS}

\end{document}